\documentclass[format=acmsmall, review=false, screen=true]{acmart}

\usepackage[utf8]{inputenc}
\usepackage{subcaption}
\usepackage{cleveref}
\usepackage{soul}
\usepackage[normalem]{ulem} 
\usepackage{booktabs} 
\usepackage{tabu}
\usepackage{multirow}
\usepackage{extarrows}
\usepackage{stmaryrd}
\usepackage{gensymb}
\usepackage{lipsum}
\usepackage{tikz}
\usepackage{xcolor}
\usepackage{float}
\usepackage{stfloats} 
\usepackage[bb=boondox]{mathalfa}
\usepackage{pifont} 
\usepackage{background}

\newcommand*\circled[1]{\tikz[baseline=(char.base)]{
            \node[shape=circle,fill,inner sep=1.2pt] (char) {\textcolor{white}{#1}};}}

\newif\ifcomment

\commenttrue

\AtBeginDocument{%
  \providecommand\BibTeX{{%
    \normalfont B\kern-0.5em{\scshape i\kern-0.25em b}\kern-0.8em\TeX}}}

\setcopyright{rightsretained}
\acmJournal{CSUR}
\acmYear{2021} \acmVolume{1} \acmNumber{1} \acmArticle{1} \acmMonth{1} \acmPrice{}\acmDOI{10.1145/3469094}

\backgroundsetup{angle=0,
    scale=1,
    color=black,
    firstpage=true,
    position=current page.north,
    hshift=0pt,
    vshift=-20pt,
    contents={\ifnum\value{page}=1 PREPRINT: Accepted for publication at the ACM Computing Surveys (CSUR) journal, 2021 \else \fi}
}

\begin{document}

\title[DNN-based Enhancement for Image \& Video Streaming Systems: 
A Survey and Future Directions]{Deep Neural Network-based Enhancement for Image and Video Streaming Systems: 
A Survey and Future Directions}

\author{Royson Lee}
\authornote{Both authors contributed equally to this research.}
\email{dsrl2@cam.ac.uk}
\orcid{0000-0002-6716-7994}
\affiliation{%
  \institution{University of Cambridge, UK}
  }
\author{Stylianos I. Venieris}
\authornotemark[1]
\email{s.venieris@samsung.com}
\affiliation{%
  \institution{Samsung AI Center, Cambridge}
}

\author{Nicholas D. Lane}
\affiliation{%
  \institution{Samsung AI Center, Cambridge \& University of Cambridge, UK}
}
\email{nic.lane@samsung.com}

\renewcommand{\shortauthors}{Lee and Venieris, et al.}

\begin{abstract}
Internet-enabled smartphones and ultra-wide displays are transforming a variety of visual apps spanning from on-demand movies and 360\textdegree~ videos to video-conferencing and live streaming. However, robustly delivering visual content under fluctuating networking conditions on devices of diverse capabilities remains an open problem. In recent years, advances in the field of deep learning on tasks such as super-resolution and image enhancement have led to unprecedented performance in generating high-quality images from low-quality ones, a process we refer to as neural \mbox{enhancement}. 
In this paper, we survey state-of-the-art content delivery systems that employ neural enhancement as a key component in achieving both fast response time and high visual quality.
We first present the components and architecture of existing content delivery systems, highlighting their challenges and motivating the use of neural enhancement models as a countermeasure.
We then cover the deployment challenges of these models and analyze existing systems and their design decisions in efficiently overcoming these technical challenges.
Additionally, we underline the key trends and common approaches across systems that target diverse use-cases. Finally, we present promising future directions based on the latest insights from deep learning research to further boost the quality of experience of content delivery systems.
\end{abstract}



\maketitle

\section{Introduction}
Internet content delivery has seen a tremendous growth over the past few years. Specifically, video traffic is estimated to account for 82\% of global Internet traffic by 2022 – up from 75\% in 2017~\cite{ciscovni,video_accel2021asplos}. This growth is attributed to not only the rapid increase of Internet-enabled devices, but also the support for higher-resolution content. For instance, an estimated 66\% of TV sets will support Ultra-High-Definition (4K) videos by 2023 as compared to 33\% in 2018~\cite{ciscoreport}. Most importantly, content traffic such as live streaming, video-conferencing, video surveillance, and both short- and long-form video-on-demand, are expected to rise very quickly. 
To meet these demands, a new class of distributed systems has emerged. 
Such systems span from content delivery systems (CDS)~\cite{Mao2017,Yeo2018} that aim to maximize the user satisfaction and quality of experience (QoE), to video analytics frameworks that provide support for augmented and virtual reality apps by co-optimizing latency and accuracy~\cite{Wang2019, Du2020}.

One of the primary challenges of distributed systems for content delivery is their reliance on networking conditions. Currently, the quality of the communication channel between client and server plays a key role in meeting the application-level performance needs due to the significant amount of transferred data and the tight latency targets. Nevertheless, in real-life mobile networks, the communication speed fluctuates and poor network conditions lead to excessive response times, dropped frames or video stalling, that rapidly degrade the user experience. This phenomenon is further amplified by the increasing number of users which compete for the same pool of network resources.

For many years, bitrate adaptation has been the dominant approach for counteracting the networking dynamicity and the main driver behind adaptive video streaming~\cite{abrsurvey2019comst}, with already deployed solutions by industry content providers such as iQiyi~\cite{abriqiyi2016sigcomm}, Netflix~\cite{abrnetflix2019mmsys} and YouTube~\cite{abryoutube2020pam}. Under this scheme, each video is broken down into segments that are encoded using multiple bitrates. Upon client's request, the segments are streamed over HTTP, with the client selecting the bitrate in a per-segment manner.
Adaptive bitrate (ABR) algorithms typically consider either the connection speed~\cite{festive2012conext,networkabr2014jsac,onrl2020mobicom} or the playback buffer state~\cite{buffer_abr2014sigcomm,bufferabr2016infocom}, to dynamically select the bitrate of each video segment.
Although ABR technology has achieved significant gains in sustaining high QoE across diverse settings, existing approaches still leave substantial portion of the available bandwidth underutilized~\cite{abryoutube2020pam,neuroabr2020jsac} either due to the ABR algorithm's suboptimality or due to the service provider's policy of minimizing bandwidth usage. With ABR leaving substantial room for further optimization, there is an emerging need for novel techniques to further boost the performance of content delivery systems and ensure high QoE across various visual content delivery applications, network conditions and device capabilities.

One of these techniques includes the use of newly emerging technologies such as telepresence~\cite{Zakharov2020}.
For instance, the recently released Nvidia Maxine~\cite{nvidia_video_conf} can be used to drastically reduce the network load by allowing each user to only transmit at least one key frame, which is used to initialize the model at the receiver's end.
During video-conferencing, facial landmarks, which use considerably less bandwidth than frames, are exchanged and used by their respective models to synthesize the video and mimic the other user's facial expressions, hence resulting in a fast and accurate video-conferencing experience even under extremely low-bandwidth settings.
However, the usage of such techniques is limited and solely for video-conferencing as working in dynamic environments or with multiple users is still an unsolved challenge.

Another recent key method that enables tackling this challenge in general is \textit{neural enhancement} through super-resolution (SR) and image enhancement models. These models are capable of processing a low-resolution or low-quality image and generating a high-quality output. With the unprecedented performance of convolutional neural networks (CNNs), content delivery systems have begun integrating neural enhancement models as a core component. The primary paradigm of using neural enhancement models in content delivery systems comprises the transmission of compact low-resolution or low-quality content, often along with the associated neural model, followed by its subsequent quality enhancement on the receiver side through a enhance-capable model~\cite{Yeo2017}. In this manner, the transfer load is minimized, drastically reducing the network footprint and the corresponding bandwidth requirements, with the visual quality recovered on the client side. 

Despite their benefits, integrating state-of-the-art neural enhancement models into visual content delivery systems poses significant challenges. First, these models, especially SR models, have excessive computational demands that are measured up to hundreds of TFLOPs per frame in order to achieve an upscaling of up to 4K or 8K.
With client platforms typically involving devices with strict resource and battery constraints~\cite{nic2017embedded_dl,venieris2018deploying,embench_2019}, clients are still struggling to execute neural enhancement models on-device while meeting the target quality~\cite{Lee2019}. This fact is aggravated by the stringent latency and throughput requirements that are imposed in order to sustain high QoE. Finally, enabling the deployment of such systems requires overcoming unique technical challenges stemming from the diversity of use-cases, spanning from on-demand video streaming~\cite{Yeo2018} to video-conferencing~\cite{Hu2019}.

In this paper, we provide a timely and up-to-date overview of the growing area of visual content delivery systems that employ neural enhancement. Through this survey, we aim to equip researchers new to the field with a comprehensive grounding of next-generation content delivery system design, revealing how neural enhancement techniques can lead to greater performance than status quo methods. More specifically, we make the following novel contributions:
\begin{itemize}
    \item We motivate neural enhancement for CDS by describing the architecture, major components, performance metrics and open challenges of conventional content delivery systems.
    \item We survey the state-of-the-art existing systems that utilize neural enhancement (Table~\ref{tab:surveyd_systems}) across diverse content delivery applications including on-demand video and image services, visual analytics, video-conferencing, live streaming and 360\textdegree~ videos. We detail their neural model design and system optimization strategies, assessing their strengths and limitations.
    \item Drawing from the latest progress by the computer vision community, we propose several promising future directions for future research and describe how they can be integrated to value-add existing and inspire future content delivery systems.
\end{itemize}

\setlength{\tabcolsep}{2pt}
\begin{table}[t]
    \centering
    \caption{Overview of Visual Content Delivery Systems}
    \resizebox{0.65\linewidth}{!}{
    \scriptsize
    \begin{tabular}{l l r c} 
        \toprule
        \begin{tabular}{@{}c@{}}\textbf{System} \\  \end{tabular} & \begin{tabular}{@{}c@{}} \textbf{Task} \\  \end{tabular} 
        & \begin{tabular}{@{}c@{}} \textbf{Year} \\  \end{tabular} 
        \\ 
        \midrule
        \texttt{MobiSR}~\cite{Lee2019} & On-demand image delivery & October 2019 \\
        Yeo \textit{et al.}~\cite{Yeo2017} & On-demand video delivery & November 2017 \\
        \texttt{NAS}~\cite{Yeo2018} & On-demand video delivery & October 2018 \\
        \texttt{NEMO}~\cite{Yeo2020} & On-demand video delivery & September 2020 \\
        \texttt{PARSEC}~\cite{Dasari2020} & 360\textdegree~ video delivery & August  2020 \\
        \texttt{Supremo}~\cite{Supremo2020} & On-demand image delivery & September 2020 \\
        \texttt{CloudSeg}~\cite{Wang2019} & Video segmentation & July 2019 \\
        \texttt{Dejavu}~\cite{Hu2019} & Video-conferencing & February 2019 \\
        \texttt{LiveNAS}~\cite{Kim2020} & Live streaming & July 2020 \\
        \texttt{SplitSR}~\cite{Liu2021} & Digital zoom & March 2021 \\
        \bottomrule
    \end{tabular}
    }
    \label{tab:surveyd_systems}
\end{table}

\section{Visual Content Delivery Systems}

\label{sec:background_vcdsys}
Content delivery systems are systems that aim to deliver image or video content to the users with minimal latency and high visual quality, while supporting a diverse set of client platforms. Across the years, several different aspects of CDS have been studied, from optimized networking~\cite{signallingbvod1997icc,harmonic1997} and system design~\cite{ge2017tmm,Mao2017,flexstream2018mm,adaptive_live2019hotedge,privatube2019middleware} to user behavior~\cite{user_vod2006eurosys,user_live2017tcsvt}, cost minimization~\cite{vod_profit2007sigcomm,li2018tpds} and performance evaluation~\cite{vod_measurements2011imc,mobile_vod_measurements2012imc,live_measurements2016sigmetrics,live_measurements2017globecom}. 
Common underlying challenges for sustaining high performance and meeting the agreed-upon quality of service (QoS) are the client device heterogeneity -- from powerful desktops to diverse mobile devices~\cite{facebook2019hpca} -- and the reliance on networking conditions.
In this section, we present \textit{1)} the common \textit{performance metrics}, \textit{2)}~the typical \textit{architecture of CDS} and \textit{3) the adaptive bitrate methods} that are conventionally used to counteract bandwidth fluctuations. Then, we introduce \textit{4)}~the recent technique of \textit{neural enhancement} in terms of its principle of operation and the associated technical challenges.

\subsection{Performance Evaluation Metrics}
\label{sec:metrics}

The performance evaluation of CDS is complex, comprising multiple metrics with often competing dynamics.
The primary metrics of interest include essential attributes such as visual quality, frame rate, response time, rebuffering time, accuracy and quality of experience. Based on the characteristics of the target task, the design of CDS often prioritizes a subset of metrics and aims to reach a task-specific balance among them. 
Therefore, measuring and reporting all these criteria across diverse scenarios plays an important role in determining the strategic trade-offs made by CDS.

\paragraph{Visual Quality}
Traditionally, the bitrate dictates the maximum resolution of the content without rebuffering; the higher the bitrate, the higher the supported resolution and thus the better the visual quality.
With the introduction of neural enhancement models in CDS, the maximum resolution is determined by both the bitrate and the amount of computational resources available to execute these models.
Additionally, the visual quality of the content is influenced not only by its maximum resolution, but also by the performance of the model used to enhance the content, \textit{i.e.}~the model's enhancement capability.
The dependency on the bitrate for visual quality is therefore relaxed and the extent of this dependency is determined by the available compute, upscale factor and degree of compression.
In other words, given sufficient computational resources, the impact on the content's visual quality is shifted towards the performance of the neural model.
For instance, higher upscale factors and more computational resources enable the streaming of high-definition resolution at lower bitrates and thus rely on the performance of the model for high visual quality. 

The visual performance of a neural enhancement model is often measured using either \textit{i)} a distortion-based metric, such as Peak Signal-to-Noise Ratio (PSNR)~\cite{gonzalez2008digital} or Structural Similarity Index Measure (SSIM)~\cite{SSIM}, or \textit{ii)} a perceptual-based metric, such as Naturalness Image Quality Evaluator (NIQE)~\cite{NIQE} or Learned Perceptual Image Patch Similarity (LPIPS)~\cite{LPIPS}.
Although optimizing using a perceptual-based metric will lead to more natural-looking images, existing CDS adopt \textit{distortion-based metrics}, which aim to maximize image fidelity.
These metrics are resolution-agnostic and the visual quality is determined solely by the absolute pixel-to-pixel error or their inter-dependencies between the original frame and its reconstructed version at the receiver.
In order to accommodate these model performance metrics into traditional QoE metrics that utilized solely the bitrates, some existing neural enhancement-based CDS directly map these model performance metrics into bitrates as detailed below.
          
\paragraph{Accuracy} 
In video analytics use-cases~\cite{Wang2019,Du2020}, such as security surveillance, traffic monitoring and face recognition, the most commonly used performance metric is accuracy, which captures the proportion of correct classifications over the input samples processed upon deployment. Accuracy is reported as percentage for classification tasks or in the range 0-1 for the Intersection-over-Union (IoU) metric of semantic segmentation tasks~\cite{long2015fully}.

\paragraph{Response Time}
\textit{Latency}, or \textit{response time}, is the primary performance indicator in latency-sensitive interactive applications, such as virtual reality~\cite{vr2018mobisys,percep_latency2015chi,vr_perf2016mm}, video analytics services~\cite{Wang2019} and video-conferencing~\cite{Hu2019}. Measured in seconds, response time is the end-to-end time between when an input event occurs (\textit{e.g.}~new frame in the user's real-time video or caller makes a gesture) and when the output is returned to the user's device (\textit{e.g.}~analytics result or enhanced view of caller). In such scenarios, excessive buffering adds a prohibitive latency overhead that degrades the QoE and is often not an option.

\paragraph{Frame Rate} Frame rate is among the primary metrics of interest in throughput-driven applications, such as video-on-demand~\cite{Yeo2018,Yeo2020,groot2020mobicom,jigsaw2019mobicom} and live video streaming~\cite{Kim2020}. Frame rate is measured in frames per second (fps), with the lower bound of real-time frame rate being between 24-30~fps.

\paragraph{Rebuffering}
A critical metric for characterizing the user-perceived quality of video applications is \textit{rebuffering}~\cite{videoqoe2011inm}. This phenomenon occurs when the playback buffer is drained due to the slow transmission of video segments, with the video player stalling and the playback pausing. 
Rebuffering captures the temporal properties of a video playback, independently of its content, and is typically analyzed with respect to: \textit{1)~initial buffering time}, the time between the request of a video by the user and the beginning of its playback; \textit{2)~mean rebuffering duration}, the average time of a rebuffering event; and \textit{3)~rebuffering frequency}, the occurrence frequency of rebuffering events.

\paragraph{Quality of Service (QoS)}
Network-level QoS captures the performance of a network connection and its capabilities to provide packet transfer with the agreed-upon requirements~\cite{qos2010network}.
QoS is typically quantified using a set of networking-level metrics, including delay, jitter, packet loss rate and bandwidth. By monitoring these quantities, service providers can tune network parameters and apply traffic shaping to meet specific service-level agreements (SLAs) and sustain a high QoS. 
In this endeavor, the expected outcome is that high network-level QoS corresponds to high user satisfaction for the underlying application.

\paragraph{Quality of Experience (QoE)}
Despite the merits of network-level QoS, the relation between QoS measurements and user satisfaction is not trivial. With a more user-centric perspective, quality of experience (QoE) aims to capture the quality of an application or service from the point of view of the users. 
QoE is typically assessed either via subjective or objective methods. To directly measure the subjective perceived quality, a group of actual users are asked to provide a quality score over a sequence of videos in order to obtain a mean opinion score (MOS)~\cite{itu2008mos}. Despite the accuracy of this approach, it relies on significant manpower and has to be conducted offline~\cite{chimp2018www}, preventing the automated monitoring of QoE that is required by real-time video applications.

With the subjective approach not deployable, objective methods have emerged. Such approaches consist of continuously collecting network-level QoS measurements together with application performance metrics~\cite{videoqoe2011inm}, such as video bitrate, mean rebuffering time, rebuffering frequency and per-frame spatial quality~\cite{videoqoe2009itng}, and using analytical formulas to aggregate them in order to automatically estimate QoE. Such an approach is adopted by the majority of the covered content delivery systems to track in real-time the achieved QoE and adapt accordingly.

The most commonly used objective QoE metric (Eq.(~\ref{eq:qoe}))~\cite{Yin2015,videoqoe2016imc,neuroabr2020jsac} to date for neural enhancement-based CDS takes into account the bitrate of each video sequence, $R$, and rebuffering measures. In this respect, QoE is generally defined as a linear combination of: the bitrate utility, $q(R)$, that maps bitrate $R$ to video quality; the rebuffering time caused by downloading the n-th segment, $T_n$; the initial buffering time, $T_s$; and the smoothness of the selected quality, $q(R_{n+1}) - q(R_n)$, which captures the variation in visual quality between subsequent segments ($n$ and $n+1$) in terms of both their number and amplitude. 
%
\begin{equation}
    \label{eq:qoe}
    QoE = \frac{\alpha \sum^N_{n=1} q(R_n) - \beta \sum^N_{n=1} T_n - \gamma \sum^{N-1}_{n=1} |q(R_{n+1}) - q(R_n)|}{N} - \delta  T_s
\end{equation}
where $N$ is the total number of video sequences, and the bitrate utility is modeled as linear~\cite{Yin2015}, $q(R) = R$, logarithmic~\cite{bufferabr2016infocom}, $q(R) = log(R/R_{\text{min}})$, to diminish improvement at higher bitrates, or fixed~\cite{Mao2017} to favor higher-definition bitrates. Hyperparameters $\alpha$, $\beta$, $\gamma$ and $\delta$ determine the importance of the four components, penalizing the average segment quality, rebuffering delay, smoothness of quality variation, and initial buffering time, respectively.

In order to take into account the neurally enhanced quality of the videos, a few works~\cite{Yeo2018,Yeo2020} used an inverse mapping from quality to bitrate through interpolation (\textit{e.g.}~piecewise linear interpolation). 
Specifically, $\widehat{R}_n = V_nf^{-1}(f(m(V_n)))$ where $V$ is the video sequence, $m$ is the neural enhancement model and $f$ is the model's performance metric. 
The QoE is, therefore, determined using the estimated bitrate, $\widehat{R}_n$, given the performance of the model instead of the actual bitrate.

\paragraph{Other considerations.}
In addition to traditional performance measures, other metrics of interest for CDS are the server-provider-side (SP-side) and user-side costs. On the SP side, costs can be quantified based on bandwidth usage, per user or aggregate, or processing time on the SP's servers. 
Analytically, these can be captured either as additional hard constraints with specific bandwidth budget and computation time limit, or as additional objectives to be optimized, where the bandwidth and server resource usage are minimized.
On the user side, cost can be quantified based on the additional energy overhead or the computational and memory resource usage imposed by running neural enhancement on the client device. Similarly to SP-side costs, these can be modeled either as hard constraints or as metrics in an optimization problem which maximizes energy efficiency and minimizes processing time.

\begin{figure}[t]
    \centering
    \includegraphics[width=0.975\textwidth,trim={0.5cm 10cm 4cm 0cm},clip]{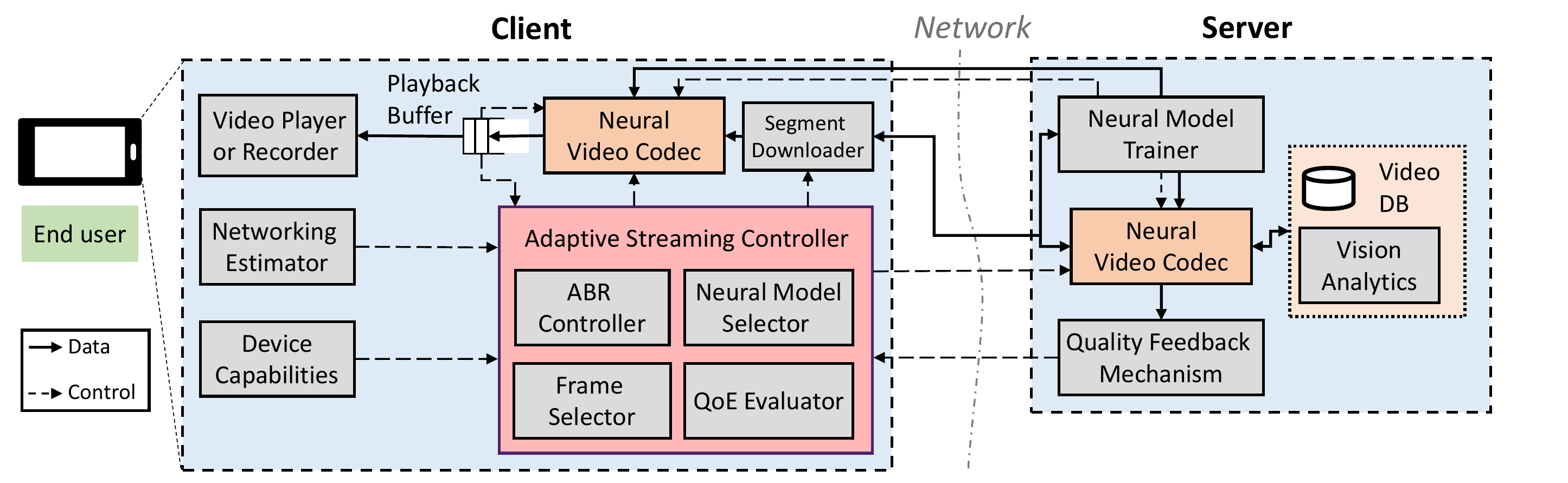}
    \caption{Architecture of content delivery systems.}
    \label{fig:content_delivery_arch}
\end{figure}

\subsection{Architecture}
\label{sec:arch}

CDS typically adopt \textit{distributed} architectures consisting of one or multiple clients and servers.
Figure~\ref{fig:content_delivery_arch} depicts the typical architecture of such a system. 
Focusing on video-on-demand (VOD) without loss of generality, the operation starts on the client side with the user selecting an online video to watch and the video player app (\textit{Video Player)} sending a request to the video server.
Next, the server fetches the video from its database (\textit{Video DB}) and launches a transmission-preparation stage. In modern adaptive video streaming services~\cite{dashstandards2011mmsys,abrsurvey2019comst}, instead of sending one large monolithic video file, the video content is divided into a series of seconds-long segments (\textit{e.g.}~4 seconds per segment). This approach enables the client's \textit{Adaptive Streaming Controller} to request different bitrate encoding per segment through the \textit{ABR Controller} in order to dynamically adapt to changing environmental conditions~\cite{elastic2013pvworkshop}. 
To this end, each video segment is compressed by the server-side \textit{Video Codec} using a predefined encoding scheme and the client-specified bitrate. 
Finally, the client's \textit{Segment Downloader} pulls the encoded segments, where they are decoded by the client-side \textit{Video Codec} and passed to the \textit{Playback Buffer} to be concatenated and eventually played. As discussed in Section~\ref{sec:landscape_vcdsys}, depending on the target application and adopted design decisions, CDS may optionally integrate additional components such as a \textit{Frame Selector}, \textit{Model Selector} and/or \textit{QoE Evaluator} on the client side, and a \textit{Quality Feedback Mechanism} and/or a \textit{Vision Analytics} backend on the server side. 
All processing stages are typically pipelined to enable streaming operations and a higher attainable throughput.

For such systems to meet the performance targets, the communication channel between client and server has to sustain high bandwidth throughout the streaming process. This constitutes a strong assumption that breaks for mobile clients where the connectivity conditions vary continuously. Hence, additional techniques, such as adaptive bitrate and neural enhancement, have been introduced that enable the dynamic adaptation to the instability of the channel. 

\subsection{Bitrate Adaptation}
\label{sec:abr}
To remedy the dependence of content delivery systems on network conditions, adaptive bitrate (ABR) algorithms have emerged~\cite{festive2012conext,networkabr2014jsac,buffer_abr2014sigcomm,bufferabr2016infocom,Yin2015,Mao2017,neuroabr2020jsac,abrsurvey2019comst,onrl2020mobicom}.
The majority of the existing bitrate adaptation methods reside on the client side, complying with the Dynamic Adaptive Streaming over HTTP (DASH) standard~\cite{dashstandards2011mmsys} and enabling large-scale deployment by applying only client-side modifications~\cite{abrsurvey2019comst}.
Under this scheme, the client device first monitors its instantaneous bandwidth (\textit{Networking Estimator} in Figure~\ref{fig:content_delivery_arch}) to assess the current network state~\cite{festive2012conext,networkabr2014jsac}, or the occupancy of its \textit{Playback Buffer}~\cite{buffer_abr2014sigcomm,bufferabr2016infocom}, or both~\cite{Yin2015,networkbufferabr2018mmsys,networkbufferabr2018sigcomm}. Next, the \textit{ABR Controller} tunes accordingly the per-segment bitrate that it requests from the server and configures the client's codec to decode at the selected rate. On the remote side, the server encodes each video segment with the specified bitrate and responds to the \textit{Segment Downloader}'s pull request. Overall, ABR techniques function as a way to control the network footprint \textit{at run time} and thus help to minimize rebuffering. 
Although ABR has been significantly improved through deep learning-based algorithms~\cite{Mao2017,deeplearningabr2017tccn,deeplearningabr2018mm,neuroabr2020jsac,onrl2020mobicom} to better perform under unexpected settings, it often fails in scarce network conditions as it relies solely on network resources. 

\subsection{Neural Enhancement}
\label{sec:neural_enhancement}

Neural enhancement aims to restore and recover the quality/resolution of visual input. As this problem is inherently ill-posed (\textit{i.e.}~many high-resolution solutions can be downsampled to the same low-resolution image), most works enforce a strong prior to mitigate its ill-posed nature. To this end, most state-of-the-art approaches utilize CNNs to capture the prior as it results in superior visual performance.
These methods train a model to either map a low- to a high-quality image using exemplar pairs~\cite{SRCNN,EDSR,VDSR} or exploit the internal recurring statistics of the image to enhance/upscale it~\cite{ZSSR}. 

The primary paradigm of using neural enhancement models in content delivery systems comprises the transmission of compact low-resolution/low-quality content followed by its subsequent enhancement on the receiver side through the enhance-capable models~\cite{Yeo2017, Yeo2018}. 
In this manner, the transfer load is tunable by means of the upscaling factor (for SR) and the degree of compression, controlling the system's network footprint and the associated bandwidth requirements. 
Therefore, neural enhancement opens up a new dimension in the design space by introducing a trade-off between computational and network resources, effectively overcoming existing systems' sole reliance on network resources.
To this end, existing systems may choose to independently optimize the utilization of these neural enhancement models~\cite{Lee2019,Kim2020,Supremo2020} or integrate them within existing ABR algorithms~\cite{Yeo2018,Dasari2020,Yeo2020}.

\section{Neural Enhancement and its Deployment Challenges}
\label{sec:dnns}

So far, a wide range of neural enhancement models and techniques have been integrated in content delivery systems.
Although the computer vision field of neural enhancement includes both SR and image enhancement neural networks, CDS generally adopt SR models as reducing the spatial dimensions is more effective at reducing network usage. 
An exception to this approach is a subset of systems~\cite{Hu2019} that repurpose SR models for image enhancement by feeding a low-quality input -- which is of the \textit{same} resolution as its output -- to the SR model.
Moreover, existing systems adopt CNNs for \textit{supervised single-image} SR, due to \textit{1)}~its efficiency as opposed to multi-frame super-resolution~\cite{EDVR,TDAN,Tao2017} and \textit{2)~the availability of the degradation operation as compared to blind} unsupervised or self-supervised methods~\cite{ZSSR, IKC, KernelGAN}. 
In this section, we provide an overview of SR models (Table~\ref{tab:dnns}) used in existing content delivery systems~(Section~\ref{sec:landscape_vcdsys}) and discuss their unique deployment challenges.

\subsection{Neural Enhancement Models used in Content Delivery Systems}

\begin{table}[t]
\caption{Comparison of neural enhancement models used in CDS (Section~\ref{sec:landscape_vcdsys}) for commonly-used scaling factors on standard benchmark datasets}
\label{tab:dnns}
    \begin{center}
    \resizebox{\textwidth}{!}{
        \begin{tabular}{lllrrcccc}
            \toprule
            \textbf{Scale} & \multicolumn{1}{c}{\textbf{Model}} & & \multicolumn{1}{c}{\begin{tabular}[c]{@{}c@{}}\textbf{Params}\\ (K)\end{tabular}} & \multicolumn{1}{c}{\begin{tabular}[c]{@{}c@{}}\textbf{Mult-Adds}\\ (G)\end{tabular}} & \multicolumn{1}{c}{\begin{tabular}[c]{@{}c@{}}\textbf{Set5~\cite{Set5}}\\ PSNR/SSIM\end{tabular}} & \multicolumn{1}{c}{\begin{tabular}[c]{@{}c@{}}\textbf{Set14~\cite{Set14}}\\ PSNR/SSIM\end{tabular}} & \multicolumn{1}{c}{\begin{tabular}[c]{@{}c@{}}\textbf{B100~\cite{B100}}\\ PSNR/SSIM\end{tabular}} & \multicolumn{1}{c}{\begin{tabular}[c]{@{}c@{}}\textbf{Urban100~\cite{Urban100}}\\
            PSNR/SSIM\end{tabular}} \\ \midrule
            
            \multicolumn{1}{c}{} & VDSR & \cite{VDSR} & 665 & 612.6 & 37.53/0.9587 & 33.03/0.9124 & 31.90/0.8960 & 30.76/0.9140 \\
            \multicolumn{1}{c}{} & CARN & \cite{CARN} & 1,592 & 222.8 & 37.76/0.9590 & 33.52/0.9166 & 32.09/0.8978 & 31.92/0.9256 \\
            \multicolumn{1}{c}{\multirow{1}{*}{2$\times$}} & IDN & \cite{IDN} & 552 & 202.8 & 37.83/0.9600 & 33.30/0.9148 & 32.08/0.8985 & 31.27/0.9196 \\
            \multicolumn{1}{c}{} & EDSR & \cite{EDSR} & 40,711 & 9384.7 & 38.11/0.9601 & 33.92/0.9195 & 32.32/0.9013 & 32.93/0.9351 \\   
            \multicolumn{1}{c}{} & MDSR & \cite{EDSR} & 7,953  & 1501.5 & 38.11/0.9602 & 33.85/0.9198 & 32.29/0.9007 & 32.84/0.9347 \\   
            \multicolumn{1}{c}{} & RCAN & \cite{RCAN} & 15,444 & 3526.8 & 38.27/0.9614 & 34.12/0.9216 & 32.41/0.9027 & 33.34/0.9384 \\ \hline
            
            \multicolumn{1}{c}{} & VDSR & \cite{VDSR} & 665 & 612.6 & 31.35/0.8838 & 28.01/0.7674 & 27.29/0.7251 & 25.18/0.7524 \\
            \multicolumn{1}{c}{} & CARN & \cite{CARN} & 1,592 & 90.9 & 32.13/0.8937 & 28.60/0.7806 & 27.58/0.7349 & 26.07/0.7837 \\
            \multicolumn{1}{c}{\multirow{1}{*}{4$\times$}} & IDN & \cite{IDN} & 552 & 89.0 & 31.82/0.890 & 28.25/0.7730 & 27.41/0.7297 & 25.41/0.7632 \\
            \multicolumn{1}{c}{} & EDSR & \cite{EDSR} & 43,070 & 2894.5 & 32.46/0.8968 & 28.80/0.7876 & 27.71/0.7420 & 26.64/0.8033 \\   
    
            \multicolumn{1}{c}{} & MDSR & \cite{EDSR} & 7,953 & 410.6 & 32.50/0.8973 & 28.72/0.7857 & 27.72/0.7418 & 26.67/0.8041 \\   
            \multicolumn{1}{c}{} & RCAN & \cite{RCAN}  & 15,592 & 916.9 & 32.63/0.9002 & 28.87/0.7889 & 27.77/0.7436 & 26.82/0.8087 \\ 
            \bottomrule
        \end{tabular}
        }
    \end{center}
\end{table}

After the first proposed super-resolution CNN~\cite{SRCNN} that only used 3 layers, VDSR~\cite{VDSR} was the first SR model that used a significantly deep (20-layer) convolutional network to obtain substantial performance gains, and is one of the first SR models adopted for on-demand video streaming~\cite{Yeo2017}. 
The model takes in an interpolated low-resolution input and passes it through twenty convolutional layers, before merging through addition the result with a final residual connection from the input.
Each convolutional layer has 64 feature maps followed by a ReLU~\cite{relu2010icml} activation function.
Although VDSR is relatively lightweight in terms of workload and memory requirements, Yeo \textit{et al.}~\cite{Yeo2017} showed that real-time upscaling to $720$p is only possible when the number of layers and feature maps are dropped by half and targeting a powerful desktop GPU, highlighting the computational challenges that will be discussed in Section~\ref{sec:challenges}.

EDSR~\cite{EDSR}, winner of the NTIRE 2017 Super-Resolution Challenge~\cite{Timofte2017NTIRE2C}, and its multi-scale variant, MDSR, are both heavily adopted in various content delivery systems such as \texttt{NAS}, \texttt{LiveNAS}, \texttt{PARSEC}, \texttt{Dejavu}, \texttt{LiveNAS} and \texttt{NEMO}.
MDSR, in particular, extends EDSR by supporting multiple scales and consists of three stages: \textit{1)} a front-end feature extraction stage with separate feature extractors per scale, \textit{2)} a number of shared intermediate layers, and \textit{3)} independent upsampling layers for each scale.
Both models adopt residual blocks from ResNet~\cite{He2016DeepRL}, without the batch normalization~\cite{BN} layers, as their building block.
Specifically, the authors showed empirically that the use of batch normalization~\cite{BN} led to performance degradation in SR, a phenomenon that was further studied in~\cite{ESRGAN}, leading to the abandonment of batch normalization layers in subsequent SR model designs.
Unlike VDSR, EDSR and MDSR employ a variety of compute-efficient techniques to reduce their computational complexity and memory footprint. 
For instance, they take in a low-resolution input and upscale it only at the end~\cite{FSRCNN} using the more efficient pixel shuffle~\cite{ESPCN} module as opposed to the more costly deconvolution.

Nevertheless, EDSR and MDSR are still considerably heavier than VDSR, with 43 million and 8 million parameters for EDSR and MDSR respectively as compared to 0.6 million parameters for VDSR. 
As a result, CDS often introduce further optimization strategies in order to alleviate the excessive workload, allow scalability based on the client's capabilities and achieve the desired performance-latency trade-off. Such techniques include the generation of multiple single-scale variants of EDSR/MDSR with fewer layers and feature maps, in order to support heterogeneous clients with varying computational capabilities~\cite{Yeo2020}.
Apart from multiple CNN configurations, early-exit strategies~\cite{branchynet2016icpr,msdnet2018iclr,hapi2020iccad,xing2020early} are often utilized to allow the client to adapt to their available resources and execute a partially downloaded model~\cite{Yeo2018}.
Other methods to speed up execution include training on the luminance channel as opposed to the RGB channels, deploying patch selection to selectively upscale patches through the models~\cite{Hu2019}, and parallelizing execution across multiple GPUs for higher-end clients~\cite{Kim2020}.

Zhang \textit{et al.}~\cite{RCAN} proposed RCAN, a model that achieves better performance at a lower cost as compared to EDSR and is adopted in \texttt{MobiSR}~\cite{Lee2019} and \texttt{SplitSR}~\cite{Liu2021}.
RCAN achieves this through channel attention blocks and a residual-in-residual structure as its building block.
Specifically, instead of designing residual blocks like in EDSR~\cite{EDSR}, RCAN consists of residual blocks within residual blocks, named residual groups. 
Each residual block in each group is followed by channel attention blocks, allowing the model to focus on the more informative components of the image.
Similar to EDSR and MDSR, RCAN is computationally heavy and requires smaller variants for it to be deployable.
For instance, apart from manually tuning the number of layers and feature maps, \texttt{MobiSR} employed a wide range of tensor decomposition and compression techniques to approximate convolutions in RCAN. Similarly, \texttt{SplitSR} substitutes the standard convolutions with channel splitting operations and tunes the depth of the RCAN to optimize for either quality or latency.

Towards efficiency, Ahn \textit{et al.}~\cite{CARN} proposed CARN which utilizes concatenated skip connections~\cite{Densenet} in a block named cascading block, which consists of residual blocks and convolutions.
The authors showed that adding concatenated skip connections at both a block-wise and layer-wise level allowed CARN to be more accurate and efficient compared to VDSR~\cite{VDSR}.
To further speed up latency, IDN~\cite{IDN} proposed channel splitting~\cite{ShuffleNetv2}, thereby only processing a subset of its feature maps for some convolutions in each block.
Moreover, IDN utilizes group convolutions~\cite{alexnet2012neurips} which can be executed in parallel.
Both CARN and IDN are computationally efficient SR models that are able to run in real-time on a high-end desktop GPU and have thus been adopted by cloud-based solutions such as \texttt{CloudSeg}~\cite{Wang2019} and \texttt{Supremo}~\cite{Supremo2020}, respectively.

Apart from the existing SR models and techniques that have already been adopted in content delivery systems, we discuss other works from the SR literature that can be utilized to further support and optimize the real-world deployment of CDS in Section~\ref{sec:future}. We refer the reader to existing surveys~\cite{Yang2019DeepLF,Wang2020DeepLF, Liu2020VideoSR} for a more comprehensive discussion on deep learning-based SR from a computer vision perspective.

\subsection{Challenges of Neural Enhancement}
\label{sec:challenges}

Despite their advantages, deep neural enhancement models pose a set of critical challenges. First, neural enhancement CNNs are extremely expensive in terms of both computational and memory burden. Key factors behind the resource-intensive nature of these models are the large spatial dimensions of feature maps throughout the model's layers together with the excessive number of memory accesses in memory-bound upscaling operations, such as Pixel Shuffle~\cite{ESPCN}.
In this respect, super-resolution CNNs are orders of magnitude larger than image discriminative models, with TFLOP-scale workloads compared to the tens of GFLOPs for classification models~\cite{embench_2019}.
Similarly, the workload of \textit{efficiency-optimized} SR models~\cite{TPSR,FSRCNN} are measured in GFLOPs, whereas their image classification counterparts~\cite{MobileNetv3} are measured in MFLOPs~\cite{embench_2019}.
A common approach to significantly reduce the peak run-time memory footprint is to split the image into patches that are processed sequentially, with the partial results stitched together to produce the final upscaled image.
Nevertheless, the computational cost is still a big challenge for real-time applications, especially when targeting mobile platforms~\cite{Lee2019,ai_benchmark_2019,Dasari2020} where the latency per frame spans from 100s of milliseconds up to seconds depending on the target image resolution.

Furthermore, models trained on standard datasets that aim to generalize across all videos/images result in outputs of varying performance upon deployment and often fail catastrophically on unexpected inputs~\cite{Yeo2017,Lee2019}. 
On the other hand, tailoring a CNN towards a specific video/image helps to mitigate this drop in performance at the cost of additional training per video/image~\cite{noscope_2017,shen2017cvpr,focus_2018}. This is typically achieved by first pretraining a model on standard datasets and then producing multiple specialized models by fine-tuning different model variants through additional training iterations either per video category~\cite{Yeo2017}, per video~\cite{Yeo2018,Yeo2020}, per video segment~\cite{Dasari2020}, per video-conference caller~\cite{Hu2019} or on a live video stream~\cite{Kim2020}.
In this respect, a \textit{generalization-specialization trade-off} is exposed which system designers need to decide how to control based on the target use-case.

\section{The Landscape of CNN-driven Visual Content Delivery}
\label{sec:landscape_vcdsys}

Despite their deployment barriers, several recent frameworks have incorporated neural enhancement methods into their pipelines and introduced novel techniques for overcoming their challenges. In this context, we survey the state-of-the-art visual content delivery systems that leverage neural enhancement models, taxonomizing them based on the type of content (on-demand or live, video or image), and provide an analysis of how they counteract \circled{a}~the excessive computational requirements and \circled{b}~the performance variability across different content. 

\subsection{On-demand Content Delivery Systems}

\subsubsection{On-demand Image Delivery}

Image delivery systems consume significantly fewer network resources than their video counterparts, therefore requiring considerably less bandwidth.
Nevertheless, apart from reducing the load in a shared communication channel, these systems can help users with a limited mobile data plan to save data.
For instance, these systems can be deployed in \textit{data-saving} mobile app alternatives, such as Facebook Lite and Messenger Lite. In this scenario, instead of disabling the downloading of content when the user is not on Wi-Fi, visual content can be downloaded in low resolution to save data, and then be locally enhanced to high quality.
In this manner, users can continue to scroll through their feed or send picture messages at ease.
This applies to any image-centric application including news apps, dating apps, gallery apps, and many others.

Additionally, as chipsets on commodity devices are gradually getting more powerful~\cite{embench_2019,ai_benchmark_2019}, this enables many of these applications to run fully on-device, avoiding the latency and privacy issues of cloud or edge offloading. 
In this direction, Lee \textit{et al.}~\cite{Lee2019} proposed \texttt{MobiSR} (Figure~\ref{fig:mobisr_arch}), a system that capitalizes over the heterogeneous compute engines of modern smartphones, \textit{e.g.} CPU, GPU and NPU, through a model selection mechanism to deliver rapid image super-resolution.

As a first step, \texttt{MobiSR} derives two model variants by applying a wide range of compression techniques on a user-provided reference model; the authors use a smaller variant of RCAN~\cite{RCAN}.
These compression techniques mainly consist of low-rank tensor decompositions, such as depthwise separable convolutions~\cite{Sifre_2014}, and other efficient model designs, such as channel splitting~\cite{ShuffleNetv2}, that have been successful in high-level vision tasks.
The resulting Pareto-optimal models are then assigned to the different available compute engines and a hardware-aware scheduler (\textit{Difficulty Evaluation Unit} (DEU)) is deployed during inference to rapidly process the patches of each input image.
Specifically, the DEU processes hard-to-upscale patches using a more compact model ($m_1$) while feeding the easier patches to a larger model ($m_2$) to obtain higher quality, leveraging on an insight that both large and small models perform similarly on hard-to-upscale patches, with difficulty quantified using the total-variation metric~\cite{total_variation_1992}.
Hence, the image quality is maximized while meeting the applications' latency constraints (challenge~\circled{a}).
With respect to challenge~\circled{b}, \texttt{MobiSR} is optimized to achieve higher overall performance through a generic model and does not employ model specialization.

\texttt{MobiSR}, however, requires scheduling using multiple processors to leverage its benefits, and may not be suitable in certain deployment scenarios.
In this regard, Liu \textit{et al.}~\cite{Liu2021} proposed \texttt{SplitSR} (Figure~\ref{fig:splitsr_arch}), a system that focuses on optimizing either quality or latency on a single CPU and, thus, making it accessible to a wider range of smartphones and computational settings.
Similar to \texttt{MobiSR}, \texttt{SplitSR} mitigates challenge~\circled{a} through an efficient variant of RCAN~\cite{RCAN}. The proposed model, however, emphasizes the use of channel splitting operations, instead of depthwise separable convolutions, and tuning the model's depth in order to reduce the DNN workload. The final DNN model is optimized for either quality or latency.
Furthermore, \texttt{SplitSR} extends the TVM deep learning compiler~\cite{tvm2018osdi} to support operations required in their model and generate highly optimized implementations for mobile CPUs.
Finally, although the models in \texttt{SplitSR} achieve higher image fidelity when compared to the models proposed in \texttt{MobiSR}, both frameworks do not specifically handle the variability in upscaling quality across images (challenge~\circled{b}).
Despite being the state-of-the-art for image-centric on-device use-cases, the processing rates achieved by both \texttt{MobiSR} and \texttt{SplitSR} are still below 24-30~fps and hence not yet suitable for real-time video applications.

\subsubsection{On-demand Video Streaming}

Video-on-demand (VOD) services allow users to watch content at their suitable time from any Internet-enabled device. The user selects a video which is then fetched and streamed by a video server to the user device. With the majority of VOD services being interactive, on-demand video streaming systems have to yield low response time and minimal rebuffering while not compromising visual quality in order to maximize the QoE. In achieving these targets, the bottleneck lies in the link between the video server and the client with the bandwidth of the connection directly affecting the end performance.

Yeo \textit{et al.}~\cite{Yeo2017} presented one of the first works that employed neural enhancement to overcome this limitation and offered a way to utilize the clients' computational power. 
Specifically, the authors first proposed grouping videos into clusters according to their category (basketball, athletics, \textit{etc.}) through a clustering module. 
This classification can be performed by either utilizing image classification models or using the video's metadata and predefined categories provided by content platform providers, such as YouTube.
A specialized SR model, VDSR~\cite{VDSR}, is then trained for each cluster, reducing the performance variation (challenge~\circled{b}) as compared to using a single generic model. 
Besides RGB frames, they also proposed the use of more compact representations such as their edges or the luminance channel, on top of tuning the spatial size of the frames, to further reduce both the bandwidth and computational resource usage.
As these alternative representations contain less information than their corresponding RGB frames, the authors utilized Generative Adversarial Network-based~\cite{Goodfellow2014} (GAN-based) training to learn the distribution of natural textures in order to synthesize natural-looking frames. 
Although the authors showed that these compact representations did not work well in practice with the H.264 codec, they were adopted in later works, such as \texttt{Dejavu}~\cite{Hu2019}, to tackle challenge~\circled{a} in different use-cases. 

To handle the excessive computational needs of neural enhancement models (challenge~\circled{a}), the authors constrained their system to work with up to 720p videos and targeted homogeneous client platforms hosting powerful desktop-grade GPUs.
This limitation was subsequently addressed to accommodate clients with heterogeneous computational capabilities in their extended proposed framework - \texttt{NAS}~\cite{Yeo2018}.

\begin{figure*}[t]
    \begin{subfigure}{.4\textwidth}
        \centering
        \includegraphics[width=1\textwidth,trim={3cm 12cm 15cm 1.8cm},clip]{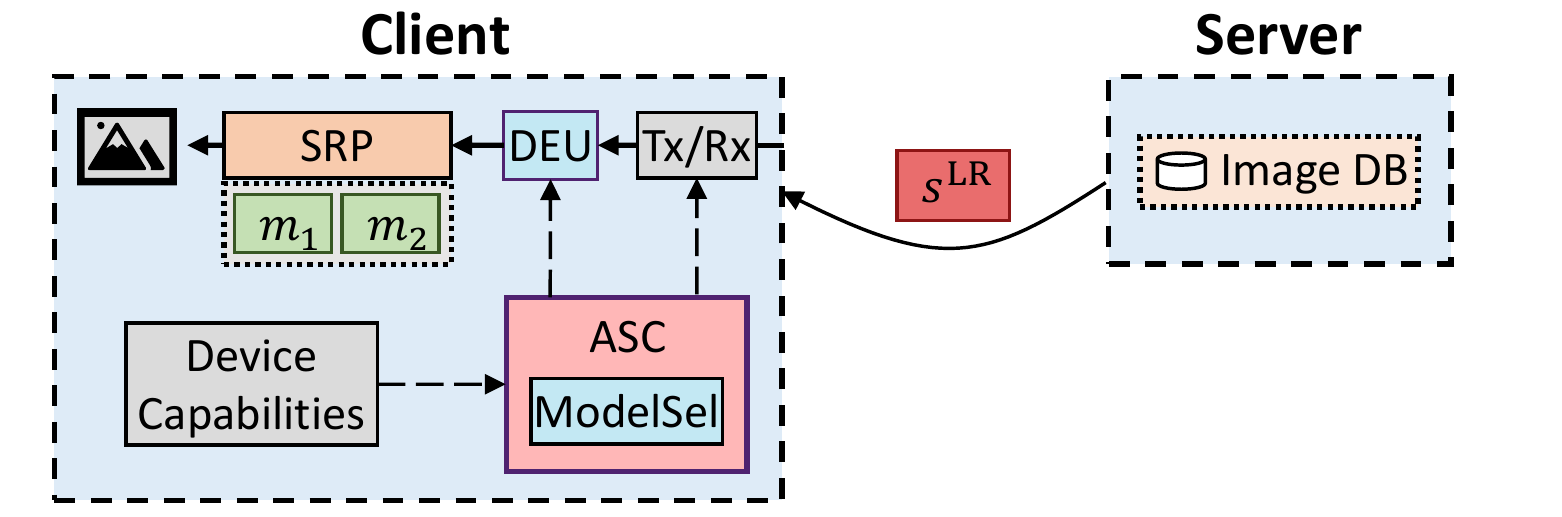}
        \caption{}
        \label{fig:mobisr_arch}
    \end{subfigure}
    \begin{subfigure}{0.4\textwidth}
        \centering
        \includegraphics[width=1.\textwidth,trim={3.5cm 11.5cm 15cm 1.5cm},clip]{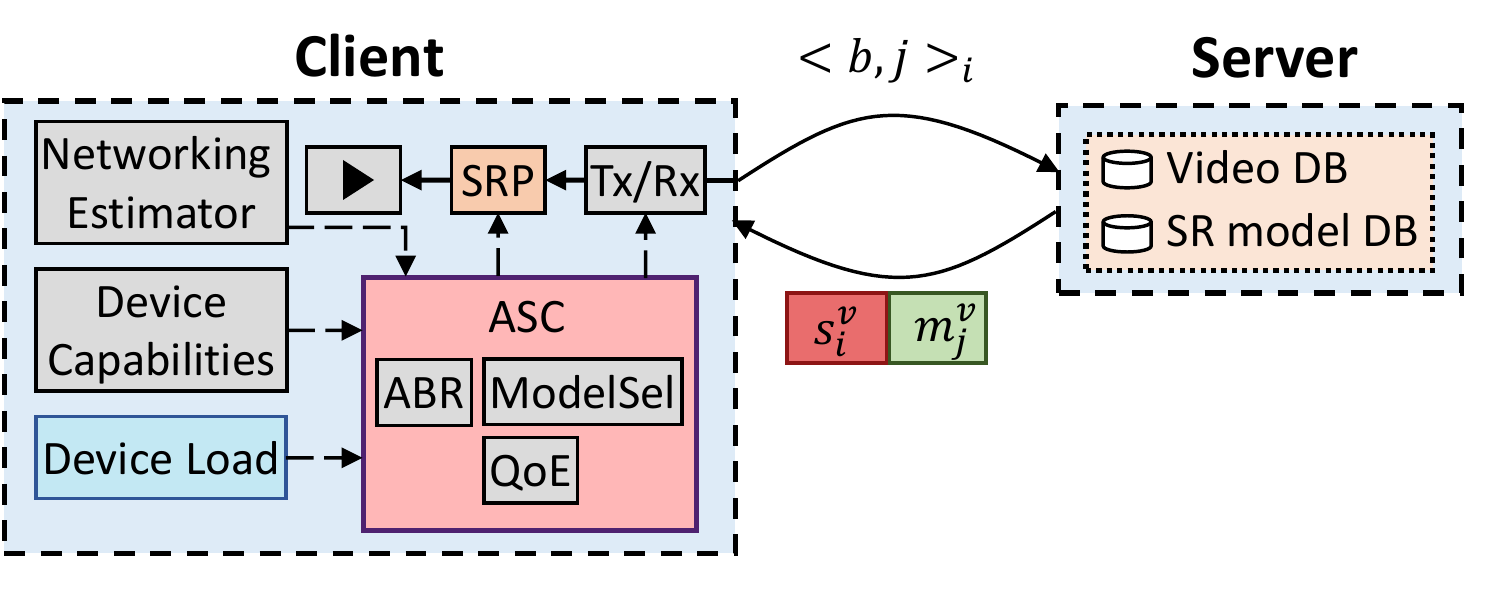}
        \vspace{-0.6cm}
        \caption{}
        \label{fig:nas_arch}
    \end{subfigure}
    \begin{subfigure}{0.4\textwidth}
        \centering
        \includegraphics[width=1.\textwidth,trim={1cm 9cm 15cm 1.5cm},clip]{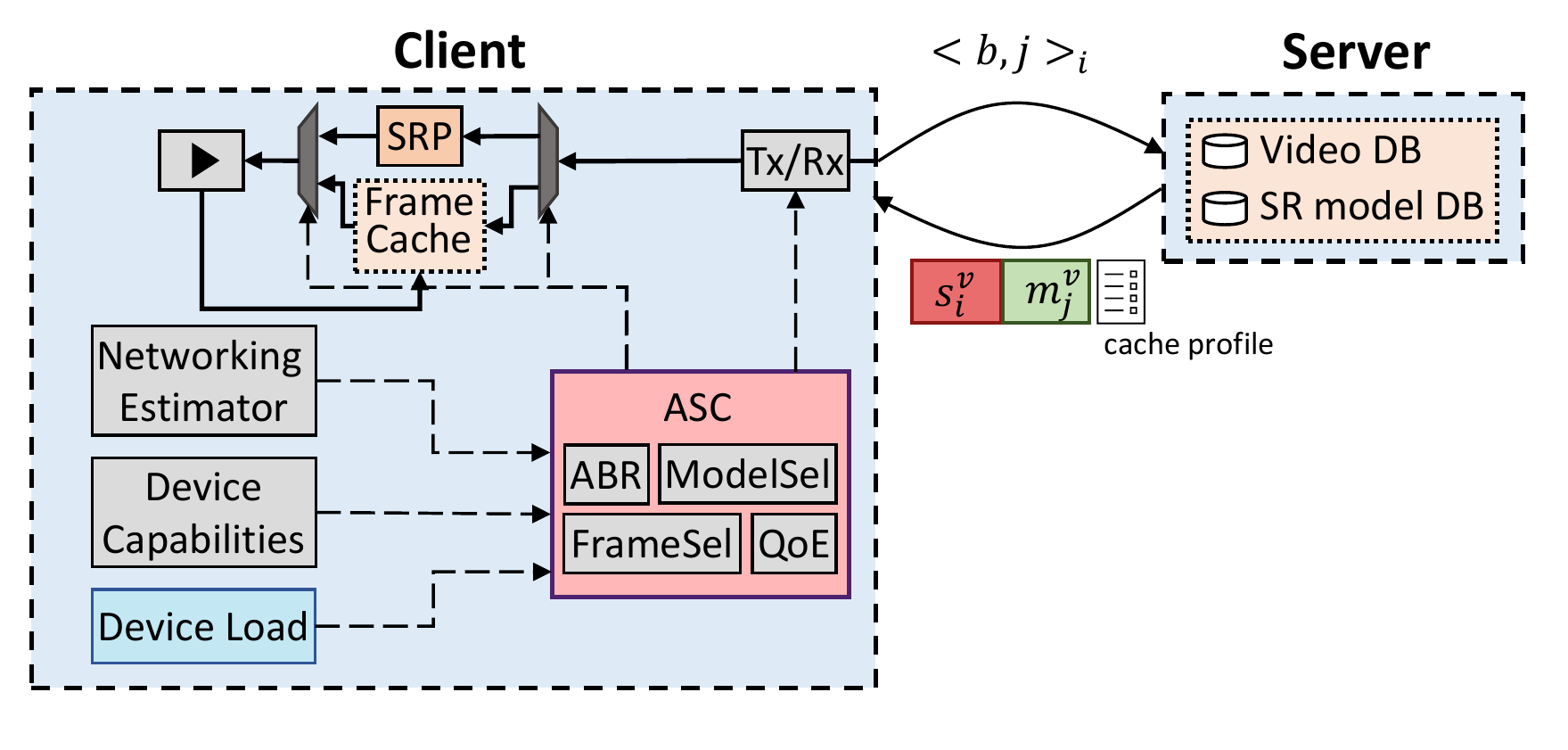}
        \caption{}
        \label{fig:nemo_arch}
    \end{subfigure}
    \begin{subfigure}{.4\textwidth}
        \centering
        \includegraphics[width=1.1\textwidth,trim={4cm 11.5cm 12.5cm 1.5cm},clip]{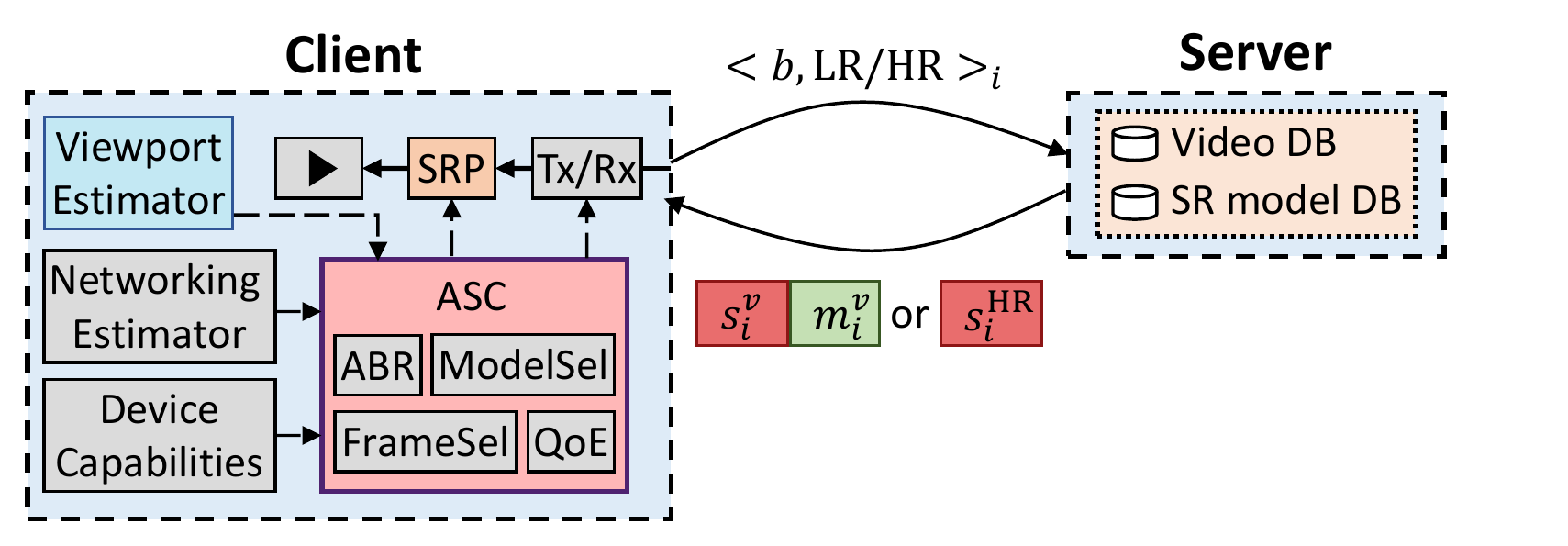}
        \vspace{0.05cm}
        \caption{}
        \label{fig:parsec_arch}
    \end{subfigure}
    \begin{subfigure}{.4\textwidth}
        \centering
        \includegraphics[width=1\textwidth,trim={2cm 11cm 15cm 1.75cm},clip]{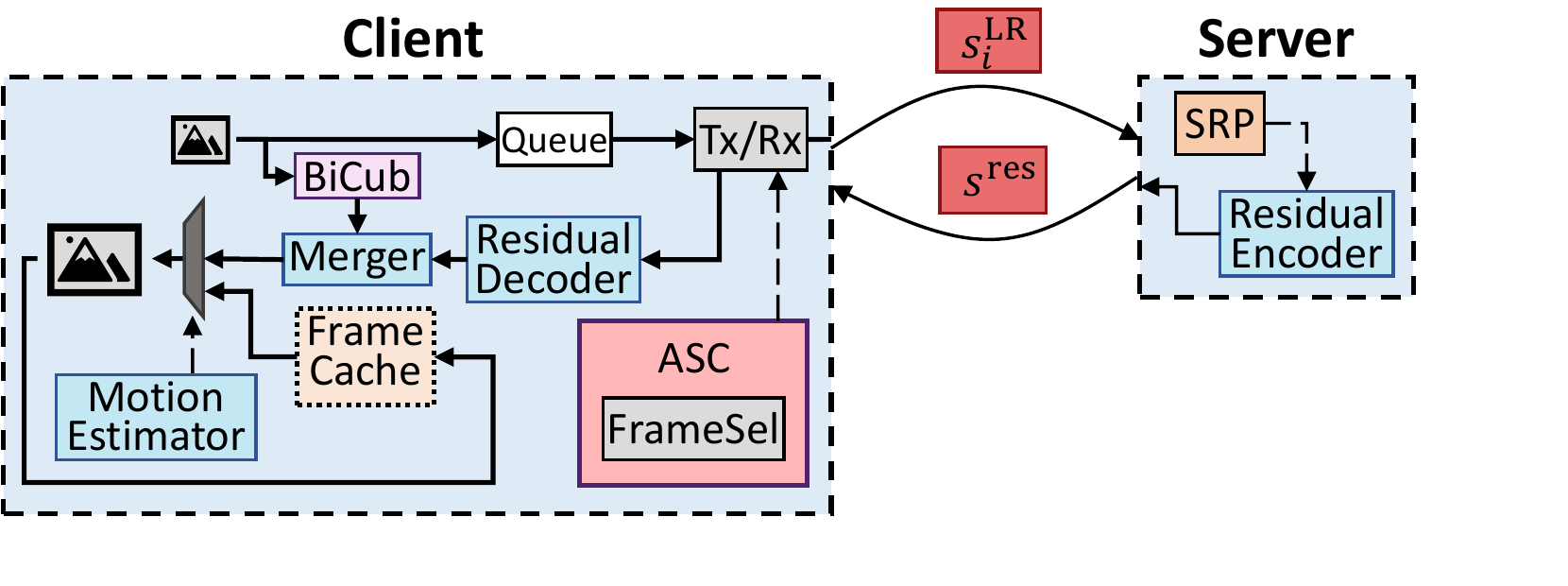}
        \caption{}
        \label{fig:supremo_arch}
    \end{subfigure}
    \begin{subfigure}{.4\textwidth}
        \centering
        \includegraphics[width=1\textwidth,trim={2cm 11cm 15cm 1.75cm},clip]{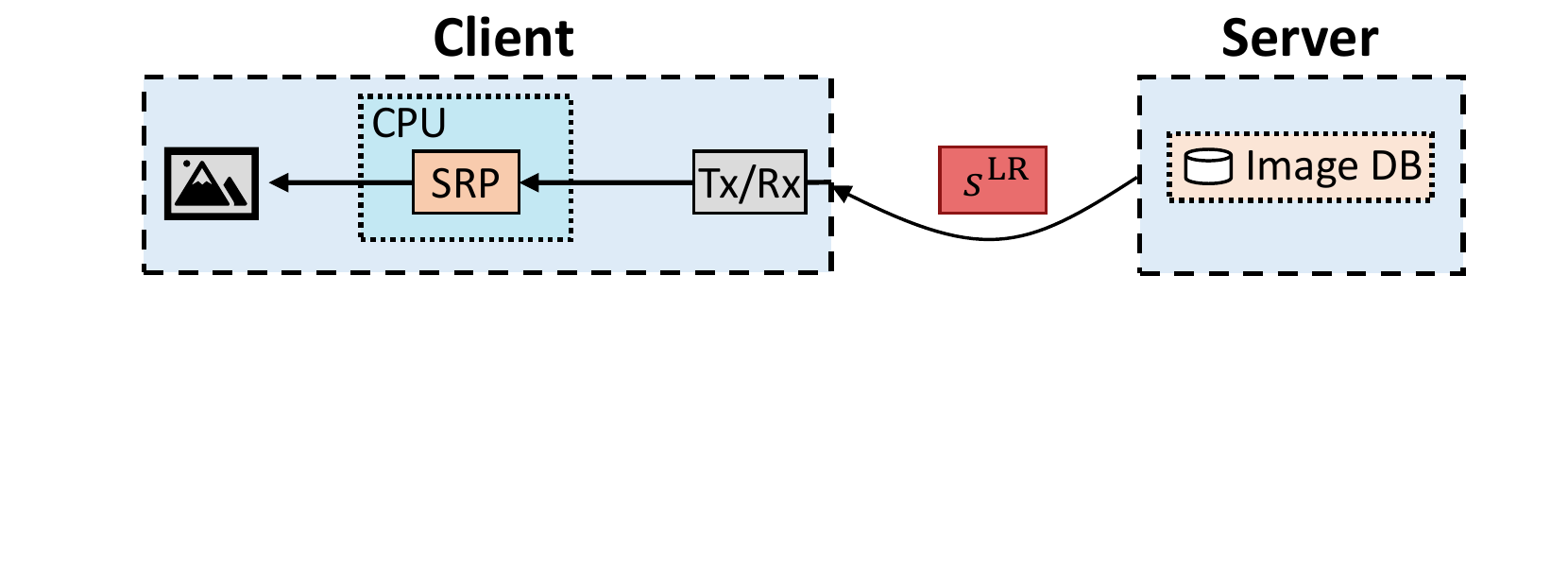}
        \caption{}
        \label{fig:splitsr_arch}
    \end{subfigure}
    \caption{Overview of on-demand delivery systems:
    \ref{fig:mobisr_arch})~\texttt{MobiSR}~\cite{Lee2019},
    \ref{fig:nas_arch})~\texttt{NAS}~\cite{Yeo2018}, 
    \ref{fig:nemo_arch})~\texttt{NEMO}~\cite{Yeo2020}, 
    \ref{fig:parsec_arch})~\texttt{PARSEC}~\cite{Dasari2020},
    \ref{fig:supremo_arch})~\texttt{Supremo}~\cite{Supremo2020}, \ref{fig:splitsr_arch})~\texttt{SplitSR}~\cite{Liu2021}. Cyan blocks indicate novel components introduced by each system.
    }
\end{figure*}

In \texttt{NAS}~\cite{Yeo2018} (Figure~\ref{fig:nas_arch}), the authors addressed the problem of heterogeneous clients through the use of early-exit SR models of varying sizes and computational workload, allowing each client to select the appropriate model segments (challenge~\circled{a}) based on both their computational capabilities and run-time device load.
To this end, they extended previous reinforcement-learning-based (RL-based) ABR algorithms~\cite{Mao2017} to decide not only the bitrate (\textit{ABR}), but also the \textit{fraction} of the SR model (\textit{Model Selector}) to be transmitted for each video segment.
Specifically, the RL-based network is optimized on a modified QoE metric to take into account the enhancement quality of the content, as detailed in Section~\ref{sec:metrics}, in order to make the aforementioned decisions.
These model segments are, therefore, progressively sent, incrementally updating the model at the client until the full model is delivered.
Further mitigating challenge~\circled{a}, the authors based their early-exit SR model on a smaller variant of MDSR~\cite{EDSR}, which is quantized at 16-bit half-precision floating-point format and executed on a desktop-grade GPU on the client side in order to hit the real-time requirement.
Finally, instead of categorizing videos into coarse clusters as in~\cite{Yeo2017}, \texttt{NAS} tackles challenge~\circled{b} by first pre-training a generic SR model and then fine-tuning a specialized model \textit{for each video}. 
Overall, as shown in Figure~\ref{fig:nas_arch}, the client selects the bitrate $b$ for the i-th video segment and the fraction $j$ of the SR model for the particular video and receives the i-th low-resolution segment $s_i^v$ and the associated fraction of the specialized model $m_j^v$ for video $v$ that are used by the \textit{Super-resolution Processor} (SRP) -- which is part of the \textit{Neural Video Codec} in Figure~\ref{fig:content_delivery_arch} -- to locally produce a high-resolution output.

Although NAS is able to support heterogeneous clients, these clients are assumed to have \textit{at least} the computational power of a desktop-class GPU.
To support lower-end commodity devices, existing on-device SR systems like \texttt{MobiSR} are inadequate at upscaling in real-time.
To this end, Yeo \textit{et al.}~\cite{Yeo2020} proposed \texttt{NEMO}, a framework that leverages frame dependencies using information from the video codec VP9, trading quality degradation for on-device real-time video streaming and reduced energy consumption. 
Instead of running SR per frame, \texttt{NEMO} applies SR to selected frames called anchor frames, mitigating challenge~\circled{a}, and uses the cached super-resolved anchor frames and the frame dependencies defined in the codec to upscale the remaining non-anchor frames. 
Specifically, the client runs a SR-integrated codec that refers to a list of anchor frames sent by the server in order to decide whether to run the model or reuse previously cached super-resolved frames for anchor and non-anchor frames respectively.
For a non-anchor frame, following the VP9 codec, the SR-integrated codec first uses a reference index to select a previously upscaled frame from the \textit{Frame Cache}. 
After this step, the provided motion vector is upscaled via bilinear interpolation and motion-compensated before it is used to warp the selected frame.
Lastly, the residual block of the compressed frame is decoded and upscaled via bilinear interpolation before adding to the resulting warped frame.

Although the VP9 codec provides key frames, which have a high-degree of reference from other frames, \texttt{NEMO} deploys its own anchor frame selection in order to adhere to a given performance threshold.
The algorithm first computes the video quality from all possible anchor point sets in a video and then, following a greedy approach, iteratively selects an anchor point that results in the maximum video quality gain until it reaches the quality requirement.
Running the quality measurements over all anchor point sets, however, is infeasible.
Therefore, in order to speed-up the algorithm, the authors selected the most impactful anchor point in each frame as a proxy, effectively reducing the number of anchor point sets to the number of frames.
The estimated video quality can thus be quickly calculated as an average of each estimated frame quality, which is computed from a single anchor point, rendering the algorithm computationally feasible.

Following \texttt{NAS}, \texttt{NEMO} alleviates challenges~\circled{a} and~\circled{b} and supports heterogeneous clients by preparing a list of models of various sizes and computational demands, along with a list of their anchor frames generated offline by its anchor frame selection algorithm, for each video and their bitrate version. 
To guarantee real-time processing, each compute unit in the client processes an anchor and a non-anchor frame once upfront in order to estimate the overall processing time required for other videos.
Given this estimation, the client then selects the available configuration (\textit{Model Selector}) that maximizes quality and meets the real-time constraint. 

\subsubsection{360\textdegree~ Video Streaming}

Compared to regular videos, streaming 360\textdegree~ videos has significantly elevated bandwidth requirements. 
To alleviate this, existing systems employ viewport prediction techniques~\cite{Fan2017} which estimate which part of the video the user will look towards and only download this spatial content. 
However, accurate viewpoint prediction is still difficult to achieve, exacerbating the problem as missing patches of the current viewpoint need to be fetched at the time of viewing.
Therefore, existing solutions~\cite{Qian2018FlarePV} send additional patches in the neighborhood of the predicted viewport patches.  
Although neural enhancement models can be utilized to mitigate this challenge, the larger spatial dimensions of 360\textdegree~ content further aggravates challenge~\circled{a} and calls for dedicated deployment solutions. 

In this context, Dasari \textit{et al.}~\cite{Dasari2020} proposed a 360\textdegree~ video streaming framework named \texttt{PARSEC} (Figure~\ref{fig:parsec_arch}). 
Unlike previous works, the authors extended the \textit{Adaptive Streaming Controller} (Figure~\ref{fig:content_delivery_arch}) logic to decide on the low-resolution (LR) patches to be upscaled~(\textit{Frame Selector}) and the bitrate of high-resolution (HR) patches to be downloaded~(\textit{ABR}) based on 1)~the networking conditions, 2) the client's computational resources, 3) the viewport prediction (\textit{Viewport Estimator}) and 4) the quality (PSNR) of both the HR and the upsampled patches. 
Since the proposed ABR algorithm is designed to maximize QoE and can thus selectively decide which patches are upscaled or downloaded, challenge~\circled{b} is mitigated.
Moreover, due to the substantial increase in spatial content of 360\textdegree~ over conventional videos, each manually tuned efficient SR model, similar to that used by \texttt{NAS}, is fine-tuned for each video segment (\textit{Model Selector}) as opposed to each video.

To overcome challenge~\circled{a}, \texttt{PARSEC} uses an extreme upscaling factor of $\times$64 on top of the H.265 compression that allows all ultra-LR patches to be transmitted, alleviating the limitations of viewpoint prediction.
Additionally, each SR model is varied depending on the targeted quality and the length of the video segment, resulting in shorter inference times for shorter video segments.

\subsubsection{Cloud-assisted Content Delivery}

In most content delivery scenarios, the client that is requesting the content adaptively allocates the compute needed for enhancement between the server and the client through its ABR algorithm as seen in \texttt{PARSEC}, \texttt{NAS}, and many others.
However, for cases in which the ABR algorithm fails, the user's QoE is heavily impacted as the client struggles to handle its workload.
In order to accommodate such cases, computation offloading can be deployed, resulting in a trade-off between performance and network utilization.
Specifically, offloading content to be enhanced offers faster processing, but imposes an extra cost in the bandwidth, along with additional privacy issues, due to the uploading and downloading of low-quality and high-quality content, respectively. Furthermore, this scheme is mainly applicable in use-cases where the low-resolution image already resides in the client device, \textit{e.g.}~pre-downloaded images or videos or zooming in camera-captured images. In such cases, there is no transmission cost to obtain the LR image and hence the additional bandwidth overhead due to offloading might be justified.

To this end, Yi \textit{et al.}~\cite{Supremo2020} proposed \texttt{Supremo} (Figure~\ref{fig:supremo_arch}), a framework that enables real-time mobile SR by selectively offloading computation to the cloud.
In order to mitigate challenge~\circled{a}, \texttt{Supremo} only runs bicubic interpolation on-device and uses a lightweight variant of the IDN~\cite{IDN} SR model to be run on the resource-rich server, while performing patch selection to only transmit key patches.
Specifically, \texttt{Supremo}'s patch selection mechanism starts by extracting the edges, using the Canny edge detector, from each image, dividing the image into blocks, and sorting the blocks according to edge intensity; the highest edge intensity has the highest priority for offloading, as edges are degraded the most when upscaled using a neural-based model compared to using plain bicubic interpolation.
Next, depending on the networking conditions, latency requirements and their ranking, these patches are sent in parallel to the cloud to be upsampled through the SR model.
To further reduce the network footprint required to download the super-resolved patches, \texttt{Supremo} exploits the sparsity of the difference between the super-resolved patches and bicubic-upsampled patches.
As these differences are often very sparse, encoding them through the \textit{Residual Encoder}, which resembles JPEG encoding, results in a heavily compressed signal, thus minimizing bandwidth. 
For video-centric use-cases, \texttt{Supremo} also deploys a caching mechanism to exploit temporal redundancy by reusing super-resolved frames on matching blocks (\textit{Frame Cache}), matched using motion estimation~\cite{Zhu2002HexagonbasedSP}.  
Similar to \texttt{MobiSR}, \texttt{Supremo} handles challenge~\circled{b} by employing a generic model that aims to maximize the average upscaling performance across all processed images.

\subsection{Live Content Streaming Systems}
\label{sec:live_content_sys}

\subsubsection{Streaming for Video Analytics}
\label{sec:video_analytics}

Pipelines for video analytics~\cite{videostorm_2017,focus_2018, Du2020,inferline2020socc} perform real-time intelligent tasks over user inputs in order to enable the development of novel applications such as telepresence~\cite{Zakharov2020} and augmented reality apps~\cite{edge_ar2019mobicom}. Such tasks span from scene labeling and object detection to face recognition. To meet the real-time performance requirements across diverse hardware platforms, such systems often rely on cloud-centric solutions. In this setup, the client device transmits the input frames to a powerful server for analysis and collects back only the result.

\begin{figure}[t]
    \begin{subfigure}{.4\textwidth}
        \centering
        \includegraphics[width=0.9\textwidth,trim={5cm 11.5cm 15cm 1.5cm},clip]{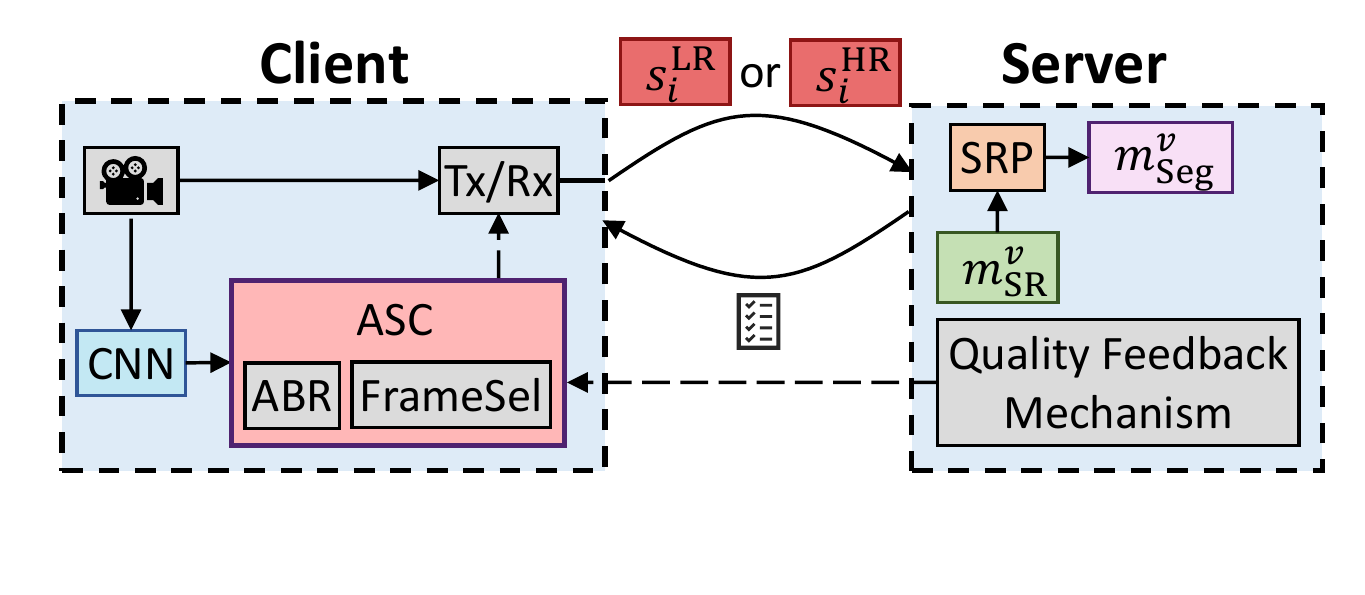}
        \vspace{1.1cm}
        \caption{}
        \label{fig:cloudseg_arch}
    \end{subfigure}
    \begin{subfigure}{.4\textwidth}
        \centering
        \includegraphics[width=1.\textwidth,trim={5cm 9cm 13cm 0.5cm},clip]{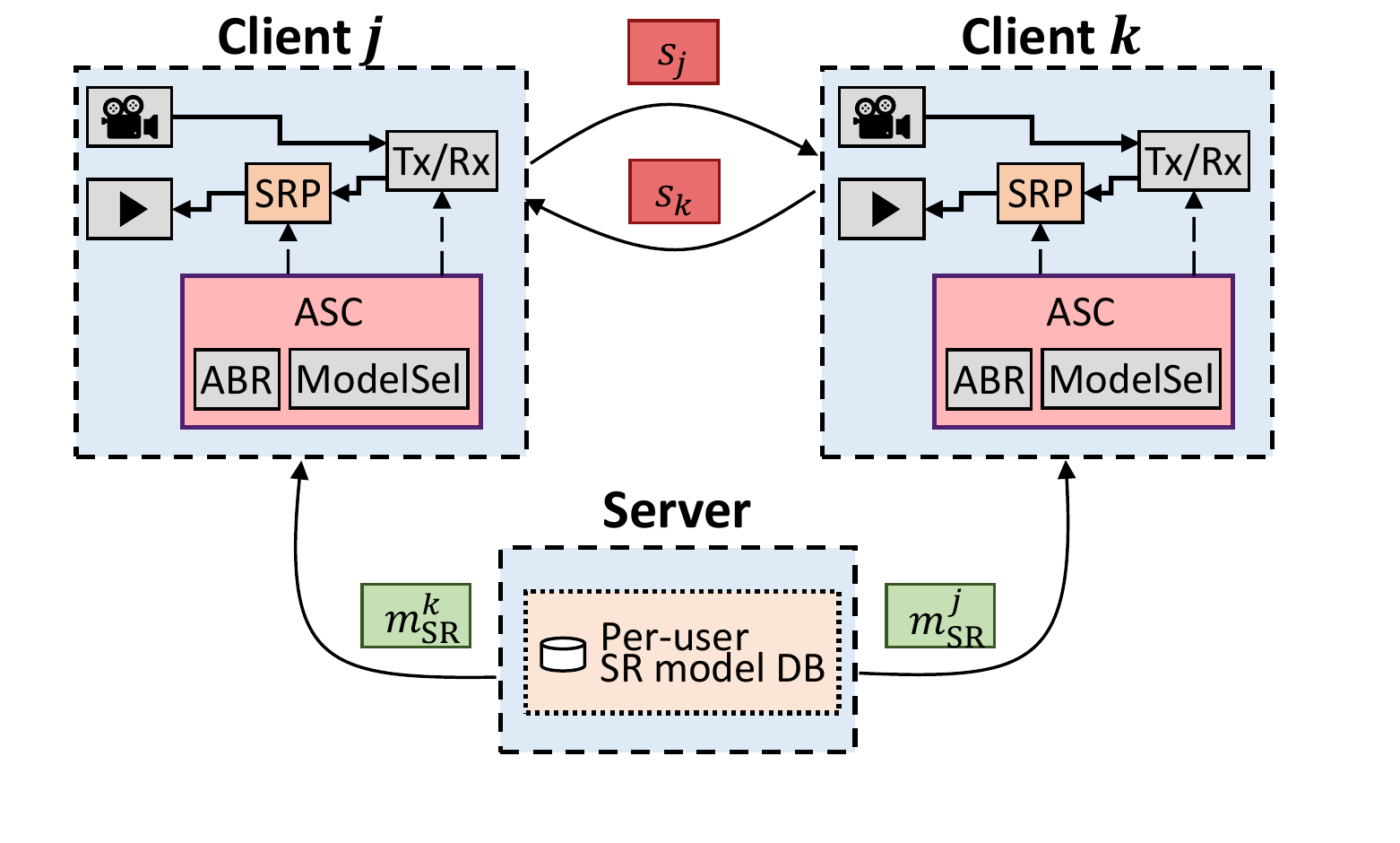}
        \caption{}
        \label{fig:dejavu_arch}
    \end{subfigure}
    \begin{subfigure}{.4\textwidth}
        \centering
        \includegraphics[width=1\textwidth,trim={3cm 9cm 15cm 1.5cm},clip]{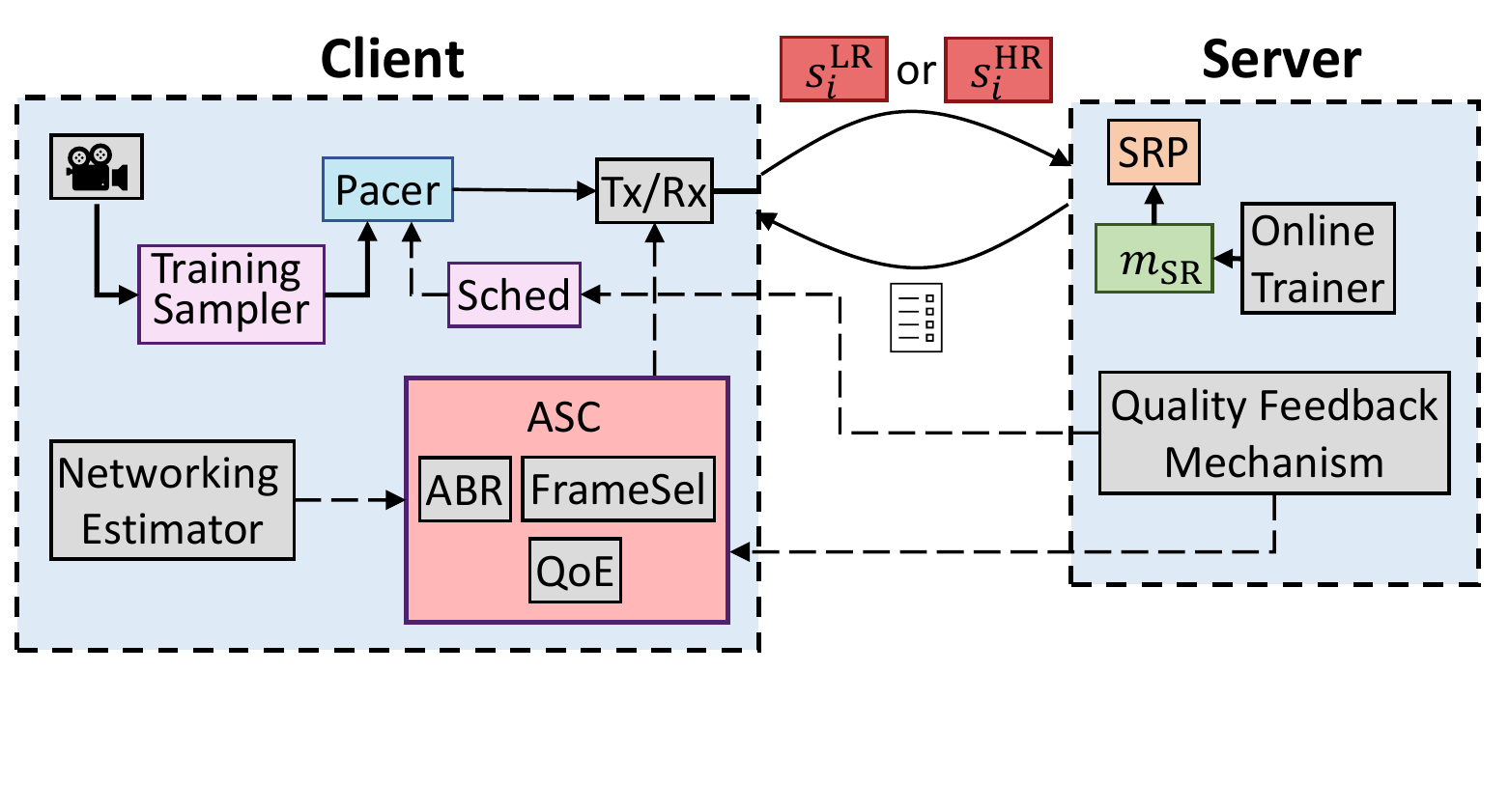}
        \vspace{-0.8cm}
        \caption{}
        \vspace{-0.2cm}
        \label{fig:livenas_arch}
    \end{subfigure}
    \caption{Overview of live streaming systems:
    \ref{fig:cloudseg_arch})~\texttt{CloudSeg}~\cite{Wang2019},
    \ref{fig:dejavu_arch})~\texttt{Dejavu}~\cite{Hu2019},
    \ref{fig:livenas_arch})~\texttt{LiveNAS}~\cite{Kim2020}. Cyan blocks indicate novel components introduced by each system.
    }
    \vspace{-0.4cm}
\end{figure}

Naturally, these video analytics frameworks can benefit from the transmission of lower-resolution or lower-quality images under bandwidth-constrained settings. 
However, the use of low-quality images is known to reduce accuracy~\cite{Chen2015,scaling2019mlsys} as some of these degraded images do not contain the sufficient information for the task at hand. 
For instance, high-frequency details, such as the texture of an object or the facial identity of distant persons, during downsampling~\cite{Cai2019ConvolutionalLF}.
Therefore, these frameworks can use additional server-side computation to deploy neural enhancement models, recover the quality of the images and thus minimize the accuracy loss of the target task. 
To achieve this goal, Wang \textit{et al.}~\cite{Wang2019} proposed \texttt{CloudSeg} (Figure~\ref{fig:cloudseg_arch}) that jointly trains an SR model (CARN~\cite{CARN}) together with its target analytics task, \textit{i.e.} semantic segmentation (ICNet~\cite{ICNet}). 
Specifically, they trained CARN using gradients computed from both the \textit{content loss} between the HR and the super-resolved image and the \textit{accuracy difference} between using both images in ICNet.

During inference, the SRP feeds both the LR image and super-resolved image to the pyramid (also known as multi-scale) segmentation model, avoiding any redundant downsampling process. 
Towards efficiency (challenge~\circled{a}), \texttt{CloudSeg} employs frame selection at the client side by deploying a small convolutional neural network (\textit{CNN}) that estimates pixel deviation in segmentation maps from low-level features of consecutive frames~\cite{Li2018LowLatencyVS} in order to skip redundant stale frames (\textit{Frame Selector)}.
Finally, \texttt{CloudSeg} overcomes challenge~\circled{b} by training both the SR and segmentation model on the same dataset and falling back to streaming in high resolution whenever the analytics accuracy falls below a threshold, \textit{i.e.}~ the \textit{Quality Feedback Mechanism} guides the \textit{ABR} controller.

\subsubsection{Video-Conferencing} 
To sustain the interactive communication between callers, video-confe-rencing requires low response time. In an effort to achieve that, existing services~\cite{Fouladi2018} often adopt conservative strategies that relax the bandwidth requirements, but also compromise visual quality. 

In this context, Hu \textit{et al.}~\cite{Hu2019} observed that in contrast to generic live streaming, video-conferencing has the property of visual similarity between \textit{recurring} sessions and designed \texttt{Dejavu} (Figure~\ref{fig:dejavu_arch}) to exploit these unique computational and specialization opportunities; video calls that take place periodically and in the same or similar rooms have inherent redundancy. 
The developed system starts with the \textit{offline} training of image enhancement models that are specialized \textit{per caller} (challenge~\circled{b}). 
In this process, the model learns to improve the video quality by increasing its \textit{encoding rate} without changing the spatial resolution. 

Upon deployment, when a video-conference session is established, the associated caller-specific enhancement model is transferred from a client to the other and vice versa ($m_{\text{SR}}^{\{k,j\}}$ in Figure~\ref{fig:dejavu_arch}). 
During the call, the frames from the caller are re-encoded to a lower quality at the same resolution prior to transmission, reducing the bandwidth usage, and the quality is enhanced on the receiver side through the caller-specific enhancement model.

To address challenge \circled{a}, \texttt{Dejavu} uses several techniques. 
Apart from designing an efficient variant of EDSR that is trained only on the luminance channel (Section~\ref{sec:dnns}), a powerful GPU is assumed to be available on each calling side. 
Additionally, \texttt{Dejavu} introduces a patch-scoring CNN that predicts the expected quality gain for each image patch. 
In this manner, only the top-$k$ patches~(\textit{Frame Selector}), which consist of mainly edges and complex textures, that are expected to yield the highest quality improvement are processed by the quality-enhancing neural network to lower the run-time resource usage.

\subsubsection{Live Video Streaming}

In contrast to VOD services that focus on stored content, live streaming targets videos produced in real-time. 
In this case, an additional bottleneck is introduced on the upstream client-to-server channel: a degraded quality from the streaming user would propagate to the end users that watch the video, leading to additional challenges in sustaining high QoE. 
Moreover, while stored content in VOD or the recurrence of video-conferencing allows for offline specialization of the enhancement models, the real-time dynamic nature of live streaming requires methods to tailor the models online to the incoming video.
Online learning, however, requires extra bandwidth on top of the stream as the parts of the high-quality stream often needs to be transmitted for model fine-tuning.

To tackle these challenges, Kim \textit{et al.}~\cite{Kim2020} proposed the \texttt{LiveNAS} system (Figure~\ref{fig:livenas_arch}) that focuses on optimizing the upstream transmission from streamer to server. 
In this system, a generic SR model, pre-trained on previously played similar streams, resides on the server side. 
Upon deployment, the streamer uploads a series of low-resolution frames which are then enhanced at the server side by the \textit{Super-resolution Processor} (\textit{SRP} in Figure~\ref{fig:livenas_arch}). 

To counteract the performance variability across diverse content (challenge~\circled{b}), \texttt{LiveNAS} introduces an online learning scheme that tailors the model to the particular unseen video.
This scheme consists of selectively picking high-quality patches using the \textit{Training Sampler} and transmitting them from the streamer to the server using the \textit{Pacer} (Figure~\ref{fig:livenas_arch}). 
Since the patches needed for online training share bandwidth with the video, it is crucial to send only the patches with the highest expected impact. 
Hence, the \textit{Training Sampler} detects patches that are hard to compress with high quality, by calculating the PSNR between HR patches and their bilinearly interpolated LR encoding, and selecting the lowest-PSNR patches.
The \textit{Pacer}, on the other hand, is responsible for allocating the available bandwidth between the low-resolution patches to be upscaled and the high-resolution training patches by adaptively tuning the respective bitrate. This is achieved through a 
scheduling algorithm (\textit{Sched} in Figure~\ref{fig:livenas_arch}), which aims to optimize the video quality together with the expected online quality gain given the total training patches thus far.

On the server side, the transferred high-resolution patches are used by the \textit{Online Trainer} to fine-tune the SR model.
Further mitigating challenge~\circled{b}, recent patches are weighted more than older patches, better reflecting the current content of the stream. 
The invocation frequency of the \textit{Online Trainer} is controlled based on an adaptive mechanism that detects training saturation and scene changes, periodically sending feedback to the client (\textit{Quality Feedback Mechanism}).
Training saturation, in particular, is tracked by measuring the performance gain from the two most recent models, suspending training and sending a signal to the client which sets the patch bitrate to a minimum value if the gain falls under a threshold consecutively.
Scene change, on the other hand, is identified by comparing the performance between the initial and latest model, resuming online training with recent patches and prompting the client to reset the patch bitrate to its initial value.
Thus, the amount of training for each live stream is adapted to maximize performance without excessive resource usage.
Finally, to alleviate challenge \circled{a}, \texttt{LiveNAS} supports scale-out execution by parallelizing the SR computations across multiple GPUs (\textit{e.g.} three GPUs for 1080p to 4K real-time enhancement), if available, on the server.

\setlength{\tabcolsep}{2pt}
\begin{table}[t]
    \centering
    \caption{Comparison of Visual Content Delivery Systems}
    \resizebox{1\linewidth}{!}{
    \scriptsize
    \begin{tabular}{l r l l c} 
        \toprule
        \begin{tabular}{@{}c@{}}\textbf{System} \\  \end{tabular} 
        & \begin{tabular}{@{}c@{}} \textbf{Ref. CNN Model} \\  \end{tabular}
        & \begin{tabular}{@{}c@{}} \textbf{Computational Optimizations} \\  \end{tabular}
        & \begin{tabular}{@{}c@{}} \textbf{Model Specialization} \\  \end{tabular}
        & \begin{tabular}{@{}l@{}} \textbf{CNN Execution}  \\  \end{tabular}
        \\ 
        \midrule
        
        \texttt{MobiSR}~\cite{Lee2019} & RCAN~\cite{RCAN} & \begin{tabular}[t]{@{}l@{}} 1) Difficulty-aware model selection \\ 2) Parallel execution \\ 3) Efficient convolutions and upsampling \end{tabular} & Generic model & Client \\
        
        Yeo \textit{et al.}~\cite{Yeo2017} & VDSR~\cite{VDSR} & \begin{tabular}[t]{@{}l@{}} 1) Powerful desktop clients \\ 2) Homogeneous clients \\ 3) Up to $720$p \end{tabular} & Offline per video category & Client \\
        
        \texttt{NAS}~\cite{Yeo2018} & MDSR~\cite{EDSR} & \begin{tabular}[t]{@{}l@{}} 1) Powerful desktop clients \\ 2) Early-exit models \\ 3) FP16 quantization \end{tabular} & Offline per video & Client \\
        
        \texttt{NEMO}~\cite{Yeo2020} & MDSR~\cite{EDSR} & \begin{tabular}[t]{@{}l@{}} 1) Key-frame selection \\ 2) Early-exit models \end{tabular} & Offline per video & Client \\
        
        \texttt{PARSEC}~\cite{Dasari2020} & MDSR~\cite{EDSR} & \begin{tabular}[t]{@{}l@{}} 1) Ultra low-resolution input \\ 2) Manually-tuned models \end{tabular} & Offline per video segment & Client \\
        
        \texttt{Supremo}~\cite{Supremo2020} & IDN~\cite{IDN} & \begin{tabular}[t]{@{}l@{}} 1) Patch selection \\ 2) Lightweight model IDN-lite \end{tabular} & Generic model & Server \\
        
        \texttt{CloudSeg}~\cite{Wang2019} & CARN~\cite{CARN} & \begin{tabular}[t]{@{}l@{}} 1) Frame selection \end{tabular} & Generic model & Server \\
        
        \texttt{Dejavu}~\cite{Hu2019} & EDSR~\cite{EDSR} & \begin{tabular}[t]{@{}l@{}} 1) Patch selection \\ 2) Luminance-only training \\ 3) Powerful desktop client \end{tabular} & Offline per caller & Client \\
        
        \texttt{LiveNAS}~\cite{Kim2020} & MDSR~\cite{EDSR} & \begin{tabular}[t]{@{}l@{}} 1) Patch Selection \\ 2) Parallel execution \end{tabular} & Online per video & Server \\
        
        \texttt{SplitSR}~\cite{Liu2021} & RCAN~\cite{RCAN} & \begin{tabular}[t]{@{}l@{}} 1) Efficient convolutions and model depth \\ 2) Compiler optimizations for CPU \end{tabular} & Quality- or latency-optimized & Client \\
        
        \bottomrule
    \end{tabular}
    }
    \label{tab:cds_comparison}
\end{table}

\subsection{Discussion: Dataset Collection}
A challenging issue of neural enhancement-based CDS systems is the collection of data for the training of the employed DNNs and the end-to-end evaluation of the whole system. Depending on whether the system is image-centric, video-centric or targets a more specialized application (\textit{e.g.}~360\textdegree videos, semantic segmentation or video-conferencing), each CDS employs a different strategy for the data collection stage. \textit{Image-centric systems}, such as \texttt{MobiSR}, \texttt{Supremo} and \texttt{SplitSR}, employ broadly used super-resolution datasets, including the DIV2K dataset for the training of their DNN models and Set5, Set14, B100 and Urban100 for the evaluation stage. \textit{Video-oriented systems}, such as \texttt{NAS}, \texttt{NEMO} and \texttt{LiveNAS}, rely on ad-hoc methodologies for the collection of datasets from existing service providers, such as YouTube. Such methodologies typically comprise a selection of the top-\textit{N} most popular videos from the top-\textit{K} most popular categories or channels on the service-provider's platform. For instance, \texttt{NEMO}, which focuses on on-demand video delivery, selected three 4K videos from the top-10 most popular categories on YouTube, while \texttt{LiveNAS}, which targets live streaming, used the most recent streams from the top streamer in Twitch's five most popular categories and one video for each of the top-4 most popular live categories on YouTube. On the other hand, \texttt{PARSEC} used the most broadly used 360\textdegree head movement dataset~\cite{360video_dataset2017mmsys} that contains head movement from 50 users on 10 360\textdegree videos from YouTube. \texttt{CloudSeg} employed the widely used Cityscapes~\cite{cityscapes2016cvpr} which is a semantic segmentation dataset with urban scenery. Finally, for \texttt{Dejavu}, the authors constructed a proprietary dataset by conducting five mock interviews under the same conditions, \textit{i.e.}~same meeting room on the same day, and recording them using a commodity smartphone on a tripod.
Regardless of the dataset used, the respective low-quality counterpart can be synthesized from the original high-resolution image using a uniform degradation operation, such as the widely used bicubic interpolation, to form input-output pairs for supervised training.
Although simple, these degradation operations suffice as the same degradation operations are used during inference. 
Nonetheless, the choice of dataset would directly affect the visual quality and the upscaling process will be considerably impacted if the data distribution encountered at run time greatly differs from the training data distribution, resulting in fail-safe mechanisms such as directly sending the HR patches~\cite{Wang2019}. To counteract this, we provide directions to improve the robustness of these models in Section~\ref{sec:future_robustness}.

\subsection{Discussion: Trends in Neural Enhancement-based Content Delivery Systems} 

State-of-the-art neural enhancement-based content delivery systems continue to face tougher deployment challenges as compared to earlier systems due to the naturally evolving landscape of various applications and devices.
These recent systems have to take into account of a wider range of heterogeneous clients and thus ensure backward compatibility with older devices.
On top of that, existing mainstream applications, such as live streaming and video calls, involve dynamic content, exacerbating challenge~\circled{b}.
Newly emerging applications such as 360\textdegree~~ videos, in contrast, have unprecedented computational demand in order to support new immersive experiences at higher resolutions (challenge~\circled{a}). 
In response to neural enhancement deployment challenges, existing CDS have deployed numerous mitigation techniques as summarized in Table~\ref{tab:cds_comparison}.
Although the diversity of target applications, with their unique requirement and challenges, does not allow us to draw conclusive and meaningful results on the performance comparison between the existing neural enhancement-based CDS, the following observations are made.

\textit{Frameworks that combine both lightweight and adaptive/scalable neural networks demonstrate higher adaptability.}
This property can be observed in CDS such as \texttt{NAS}, \texttt{NEMO}, \texttt{MobiSR} and \texttt{SplitSR}. On the one hand, \texttt{NAS} and \texttt{NEMO} employ: \textit{1)} a lightweight model design through a compact variant of MDSR~\cite{EDSR} with 16-bit quantization, together with \textit{2.1)}~a per-video catalog of models with diverse complexity-performance trade-offs that enables the \textit{static} model selection based on the client device nominal capabilities, and \textit{2.2)} a scalable design of individual models that enables the \textit{dynamic} adaptation of model complexity to the run-time device load.
Adopting an alternative approach, \texttt{MobiSR} combines: \textit{1)} a lightweight compressed version of RCAN~\cite{RCAN}, together with \textit{2.1)} a \textit{static} model selection based on the target device capabilities and model performance and \textit{2.2)} a \textit{run-time} model selection scheduler that pins a distinct model to each compute engine of the target platform and dispatches each patch to the suitable model-engine pair. Similarly, \texttt{SplitSR} employs a customized version of RCAN, optimized for either quality or latency.
Such approaches provide greater flexibility at both design and run time. In this manner, CDS have more room to adapt to static (\textit{e.g.}~device capabilities) and dynamic constraints (\textit{e.g.}~device load, networking conditions), meet the latency and visual quality requirements, and maximize QoE.

\textit{Applying informed frame/patch-selective enhancement through a quality-aware mechanism tends to radically improve QoE with minimal visual quality degradation.}
\texttt{NEMO} and \texttt{Supremo}'s caches that store previously super-resolved frames enables the reuse of already computed results, thus skipping the costly processing overhead of neural enhancement. With the goal to sustain high visual quality in the challenging task of 360\textdegree~~ video delivery, \texttt{PARSEC} selectively chooses which frame patches to enhance and which to directly download in high resolution. A similar strategy is adopted by \texttt{CloudSeg} with the aim to maximize the accuracy of the target semantic segmentation task. In another direction, \texttt{LiveNAS} employs frame selection to determine which high-resolution frames should be transmitted to the server side for further neural model fine-tuning on the current live stream.
Across these CDS, the frame/patch selection criteria aim to either select content based on their upscaling or encoding difficulty or the degree of temporal redundancy. For instance, patches with complex textures and edges are often prioritized to be enhanced in order to maximize performance, or frames with high correlations with subsequent frames are cached after enhancement in order to be reused.

\textit{Specializing models to specific content tends to reduce the complexity of neural model design and rely more on system-level solutions.}
General mitigations to challenge~\circled{b} involve fine-tuning from a pre-trained generic model to the targeted content per scale for each client's computational regime.
With the exemption of \texttt{MobiSR}, \texttt{Supremo}, and \texttt{CloudSeg}, all existing CDS adopt a variant of this model specialization approach.
These systems require $C*S*\epsilon$ trained models where $C$ is the number of unique content (\textit{i.e.}~entire video, video cluster or video segment), $S$ is the number of available scales for each content, and $\epsilon$ is the number of supported computational regimes across its clients (\textit{e.g.}~low-end, mid-tier and high-end platforms).
This computational load is a one-time cost for offline applications, but imposes additional challenges for their online counterparts, such as \texttt{LiveNAS}, as discussed in Section~\ref{sec:live_content_sys}.
This approach relieves CDS designers from designing sophisticated neural enhancement models by deriving multiple specialized and scale-specific variants from a single reference model, with the CDS deploying the suitable one as requested by the client. Nevertheless, it leaves space for further model-level exploration as discussed in Section~\ref{sec:efficient_models}.

\section{Future Directions}
\label{sec:future}

In this section, we propose various approaches that are drawn from the latest computer vision research and provide insights on how neural enhancement can further benefit content delivery systems. 
Specifically, we highlight promising directions to further address the challenges of neural enhancement.

\subsection{Improving Visual Quality}

One of the main open challenges in neural enhancement algorithms is the design of a metric that will correspond well with human raters. 
As mentioned in Section~\ref{sec:metrics}, distortion-based metrics, such as PSNR~\cite{gonzalez2008digital} or SSIM~\cite{SSIM}, which aim to minimize the per-pixel error between two images, 
have been extensively shown to improve image fidelity at the cost of perceptual quality, leading to blurry and unnatural outcomes~\cite{SRGAN}.
On the other hand, optimizing only for a perceptual-based metric such as Naturalness Image Quality Evaluator (NIQE)~\cite{NIQE} and Learned Perceptual Image Patch Similarity (LPIPS)~\cite{LPIPS} will lead to more natural-looking images at the expense of fidelity and therefore the occasional occurrence of image artefacts. 
Mathematically, there is a trade-off between fidelity and perceptual quality~\cite{Blau_2018}.

As all the existing neural enhancement frameworks (Section~\ref{sec:landscape_vcdsys}) train their models using a distortion-based metric, the outputs of these models are accurate, but may look unnatural. 
Although this will benefit video analytics frameworks, such as \texttt{CloudSeg}, having blurry outputs will undermine the goal of other content streaming systems, such as \texttt{Dejavu} and \texttt{LiveNAS}. 
To close this gap, these systems can benefit further by utilizing recent methods proposed in computer vision to train and optimize their neural enhancement models. 
Some of these works focus on striking an optimal point between image fidelity and perceptual quality by optimizing for both distortion-based and perceptual-based metrics~\cite{ESRGAN,TPSR}.
Specifically, these works often optimize their models jointly on L1/Mean-Squared-Error (MSE) losses for image fidelity and a variety of losses, including perceptual loss~\cite{Johnson2016}, adversarial loss~\cite{Goodfellow2014}, and contextual loss~\cite{Mechrez2018MaintainingNI}, for perceptual quality.
Additionally, other methods such as interpolating between a distortion-based and perceptual-based output image~\cite{ImageInterpolation} or model~\cite{ESRGAN,NetworkInterpolation}, or introducing additional priors~\cite{SPSR,NatSR}, can also be used to alleviate undesirable artefacts caused by optimizing only for perception-based metrics.
For instance, utilizing rich texture priors found in pre-trained generative models has been shown to result in images that are high in both fidelity and perceptual quality~\cite{GLEAN}.
Although the aforementioned works have found success in achieving high visual quality content, most propose techniques and models that are too computationally demanding to be deployed in existing CDS. Hence, there are potential gaps in the literature to realize the benefits of utilizing these approaches.

For video-based solutions, the majority of existing vision works utilize temporal information by aligning and fusing spatial information from multiple frames to further boost image fidelity~\cite{Liao2015, Kappeler2016, FRVSR, BASICVSR}.
Specifically, these CNN- or RNN-based solutions align adjacent frames by warping each supporting frame to a reference frame using the respective optical flow.
These optical flows are usually estimated using either traditional motion estimation algorithms~\cite{Baker2007} or CNN-based approaches, such as spatial transformer networks~\cite{Jaderberg2015} and task-specific motion estimation networks~\cite{FlowNet2, SPynet, PWCNet}.
Instead of utilizing an explicit component for motion estimation, another line of work~\cite{Jo2018, TDAN, EDVR} has proposed performing alignment implicitly through deformable convolutions~\cite{Dai2017, Zhu2019} or dynamically-generated filters.

As a result of effectively utilizing multiple frames to upscale each frame, these multi-frame approaches are able to achieve better restoration performance as compared to their single-image counterparts.
Besides improving the visual quality spatially, temporal interpolation methods~\cite{Zuckerman2020,Xiang2020} can enable a system to increase the achieved frame rate, and hence the QoE, by estimating intermediate frames rather than transferring them and applying super-resolution. 
In this manner, the bandwidth requirements are reduced by cutting both the content's spatial and temporal resolution.
Nonetheless, video-based enhancement solutions are more computationally demanding than image-based solutions due to additional processing in the temporal domain. Therefore, these techniques can be utilized more effectively in CDS such as \texttt{NAS} and \texttt{CloudSeg}, which perform their computations on powerful desktop clients and on server-grade processors, respectively.

With regards to QoE, the actual bitrate or an estimated bitrate, given the performance of the model, is used to quantify the content's visual quality in relations to the user's experience (Section~\ref{sec:metrics}).
As the amount of computational resources continue to grow in commodity hardware, the content's spatial quality will be largely dependent on the performance of the model.
Therefore, there is a need for better designs of the QoE metric to reflect these changes for neural enhancement-based CDS.

\subsection{Utilizing Efficiency-optimized Models} 
\label{sec:efficient_models}
Most neural enhancement frameworks generally adopt popular full-blown SR models (Table~\ref{tab:cds_comparison}) and revise them in order to speed up training and inference or fit into the limiting constraints for client-side computation. However, these revisions are usually suboptimal or, in some cases, detrimental. 
For instance, \texttt{PARSEC} uses of batch normalization~\cite{BN} to speed up training, reducing image fidelity~\cite{EDSR} and introducing image artefacts~\cite{ESRGAN}. 
Therefore, instead of naively scaling down and modifying full-blown SR models, these systems can leverage existing off-the-shelf efficient models that are specifically optimized for both performance and efficiency in order to deliver higher-quality enhancement at a lower computational cost (challenge~\circled{a}).
These models include manually designed variants such as IDN~\cite{IDN} -- already used by \texttt{Supremo}, automatically-designed variants such as ESRN~\cite{ESRN} and TPSR~\cite{TPSR} through neural architecture search~\cite{Zoph2017NeuralAS}, or even quantized~\cite{PAMS2020eccv} and binarized SR models~\cite{Ma2019, Xin2020} if the hardware supports their efficient execution.

\subsection{Faster Model Specialization through Meta-learning}
To mitigate challenge~\circled{b}, some systems specialize a model for each specific image/video through overfitting. This approach results in a computationally expensive process (challenge ~\circled{a}) as each model requires thousands of gradient updates in order to be adequately specialized.
In response, later systems introduce a two-stage approach. First, a generic neural enhancement model is pre-trained offline, and then fine-tuning is applied either offline or online.
For instance, \texttt{NAS} first trains a generic model on an external dataset before using its weights to fine-tune a separate model for each video, amortizing in this manner the one-time offline training cost. 

To further speed up and improve the performance during the fine-tuning step, these works can adopt a \textit{meta-learning} approach~\cite{hospedales2020metalearning} in order to find a more optimal set of initialization parameters for fine-tuning. 
In other words, pre-training a neural enhancement model via meta-learning on an external dataset will require fewer gradient updates during the fine-tuning stage, therefore requiring fewer computational resources and leading to higher performance compared to brute-force fine-tuning~\cite{Soh2020,Park2020}.

This fast adaptation is achieved by training the model so that it is easy to fine-tune.
Specifically, besides having a regular training loop that optimizes the model on an external dataset, there is an additional inner loop that computes the adapted parameters given a different set of paired images.
The main set of parameters is then iteratively updated based on the derivative of the loss with respect to the adapted parameters, resulting in a set of parameters that can quickly adapt to a test image during inference.
Therefore, the quick adaption of the meta-trained parameters can further save the computations required to fine-tune in both offline (\textit{e.g.}~on-demand streaming) and online cases (\textit{e.g.}~live streaming).

\subsection{Improving Robustness through Conditional Enhancement}
\label{sec:future_robustness}
Originally, the usage of SR was to improve the quality of a low-quality image under the assumption that the high-quality version was not available.
As such, one of the key benefits of deploying an SR model as a neural enhancement unit is its ability to work without the high-resolution \textit{ground-truth}. 
However, in the context of many content delivery settings, the ground-truth \textit{is} available and the service provider scales it down to a lower quality in order to minimize the transmission cost.
Therefore, instead of discarding the high-frequency information during downsampling, this information can be encoded in a latent variable and transmitted, along with the low-quality frame, to the receiver in CDS.
The upsampling process can then be conditioned on that variable, therefore counteracting performance variability (challenge~\circled{b}) by allowing the same model to better handle differing content instead of having to overfit one model for each content.
To this end, several computer vision works leverage the downscaling process during the upscaling process using neural image rescaling techniques to further boost image reconstruction. 
For instance, a downscaling CNN can be trained jointly with an existing SR model as shown in~\cite{Kim2018} and techniques such as encoder-decoder frameworks and invertible neural networks can also be utilized 
as shown in \cite{Li2019} and \cite{Xiao2020} respectively. 

Despite its benefits, neural-based image rescaling incurs an additional cost of executing a downscaling neural network as compared to that of interpolation methods, utilizing additional computational resources for a more robust improvement in visual quality. 
Therefore, image rescaling techniques may be more suitable for on-demand video systems, such as \texttt{NAS} and \texttt{PARSEC}, in which the downscaling cost is an offline one-time cost across videos. 
Additionally, although these approaches have not been shown to fully mitigate challenge~\circled{b}, they can significantly reduce the dependency of having to frequently transmit model segments for each content.
Ultimately, as models get more robust to varying inputs, they can be deployed once across varying content and even across various applications.

\subsection{Dynamic Deep Neural Networks}
\label{sec:dynamic_dnns}

A growing body of work in the computer vision literature is investigating the design and deployment of dynamic deep neural networks. Such models employ conditional execution mechanisms in order to provide a run-time tunable accuracy-complexity trade-off, by scaling up and down their computational complexity.
Such mechanisms would allow CDS to dynamically adapt the execution of neural enhancement models based on the client device capabilities and compute load, and provide another configurable dimension for CDS to ensure high QoE.
Nevertheless, the majority of existing work focuses on classification models~\cite{wang2018skipnet,nestdnn2018mobicom,gao2019dynamic,gatingnns2019neurips,hapi2020iccad,see2019mswim,spinn2020mobicom,branchynet2016icpr,sdn2019icml,mcdnn_2016,noscope_2017,focus_2018,shen2017cvpr,cascadecnn2018,cascadecnn2020date}, rather than super-resolution and image enhancement.

Although \texttt{NAS}~\cite{Yeo2018} and \texttt{MobiSR}~\cite{Lee2019} explored this direction by introducing the scalable NAS-MDSR model and a difficulty-aware model selection scheme, respectively, the potential of other dynamic methods has remained unexplored. So far, the computer vision community has proposed a broader range of dynamic neural network techniques, spanning from \textit{dynamic pruning} schemes that skip either layers~\cite{wang2018skipnet} or channels~\cite{nestdnn2018mobicom,gao2019dynamic,gatingnns2019neurips,wang2020dynamic} in an input-dependent manner, \textit{model cascades}~\cite{mcdnn_2016,noscope_2017,focus_2018,shen2017cvpr,cascadecnn2018,cascadecnn2020date}, and \textit{early-exit models} (either hand-crafted~\cite{msdnet2018iclr,scan2019neurips}, hardware-aware~\cite{hapi2020iccad}, distributed~\cite{see2019mswim,spinn2020mobicom} or generic~\cite{branchynet2016icpr,sdn2019icml}). By adapting and modifying such techniques, system designers can develop dynamic neural enhancement models~\cite{xing2020early} that can be highly optimized specifically for the deployment use-cases of content delivery systems.

\subsection{Hardware Acceleration through Neural Processing Units}
\label{sec:hw_accel}

Another promising approach to alleviate the high computational demands of neural enhancement models is targeting the problem from a hardware perspective. At the moment, device vendors have integrated specialized hardware units -- often named \textit{neural processing units} -- that are optimized for fast deep neural network processing in both smartphones~\cite{ai_benchmark_2019} and embedded platforms~\cite{embench_2019}. To surpass the limitations of CPUs and GPUs and further boost the processing speed of neural models, several works have proposed custom hardware accelerators~\cite{sv2019tnnls,samsungnpu2019isscc}.

In this context, by considering the unique needs of content delivery systems and their use-cases, highly customized hardware accelerators can be designed. This approach would tailor the underlying processing platform to the CDS' performance requirements, overcoming the performance and energy-efficiency bottlenecks of conventional CPUs and GPUs, and thus ensure high QoE at a lower overall cost. 
Such custom hardware solutions can be obtained either through hand-crafted designs~\cite{He2018fccm} and super-resolution neural processing units~\cite{srnpu2020jecas}, automated accelerator generation tools~\cite{sv2018csur}, or sophisticated model-hardware co-design frameworks~\cite{fpga_sr_2018tcsvt,abdelfattah2020best,Latbench} for the joint development of the neural enhancement model and its efficient hardware accelerator.

\section{Conclusion}
\label{sec:conclusion}

This paper presents a survey of a new class of visual content delivery systems that employ neural enhancement techniques to boost their performance. 
Through a detailed analysis of how they tackle the deployment challenges of deep neural enhancement models, we highlight their design choices in terms of system architecture, novel components and neural model design, and indicate their strengths and weaknesses.
Despite the rapid progress of recent systems, the demand for visual content traffic will grow over the coming years due to both existing and emerging technologies, such as augmented and virtual reality~\cite{overlay2015mobisys,edge_ar2019mobicom} and telepresence~\cite{Zakharov2020}. 
To this end, based on recent developments from both the computer vision and systems communities, we identify key optimization opportunities and propose promising research directions to address emerging challenges of the field, enhancing their performance and enabling a wider large-scale deployment of high-QoE content delivery services.


\bibliographystyle{ACM-Reference-Format}
\bibliography{ref}

\end{document}